\documentclass[airedicted sol,jcp,amsmath,amssymb,floatfix,reprint,citeautoscript,noeprint,longbibliography]{revtex4-1}

\usepackage[utf8]{inputenc}
\usepackage{graphicx}
\usepackage{wrapfig}
\usepackage[colorlinks,allcolors=black,citecolor=blue,urlcolor=blue]{hyperref}
\usepackage[mathlines,displaymath]{lineno}
\usepackage{chemformula}
\usepackage{placeins}

\DeclareUnicodeCharacter{2212}{-}

\graphicspath{{figures/}}

\begin{document}

\def\mytitle{%
Crumbling Crystals: On the Dissolution Mechanism of NaCl in Water
}
\title{\mytitle}

\author{Niamh O'Neill}%
\affiliation{%
Yusuf Hamied Department of Chemistry, University of Cambridge, Lensfield Road, Cambridge, CB2 1EW, UK
}

\author{Christoph Schran}%
\email{cs2121@cam.ac.uk}
\affiliation{%
Yusuf Hamied Department of Chemistry, University of Cambridge, Lensfield Road, Cambridge, CB2 1EW, UK
}
\affiliation{%
Thomas Young Centre, London Centre for Nanotechnology, and Department of Physics and Astronomy, University College London, Gower Street, London, WC1E 6BT, United Kingdom
}%
\author{Stephen J. Cox}%
\affiliation{%
Yusuf Hamied Department of Chemistry, University of Cambridge, Lensfield Road, Cambridge, CB2 1EW, UK
}

\author{Angelos Michaelides}%
\email{am452@cam.ac.uk}
\affiliation{%
Yusuf Hamied Department of Chemistry, University of Cambridge, Lensfield Road, Cambridge, CB2 1EW, UK
}
\affiliation{%
Thomas Young Centre, London Centre for Nanotechnology, and Department of Physics and Astronomy, University College London, Gower Street, London, WC1E 6BT, United Kingdom
}%

\keywords{Dissolution, Aqueous Phase, Molecular Simulations}

\begin{abstract}
Life on Earth depends upon the dissolution of ionic salts in water, particularly NaCl. 
However, an atomistic scale understanding of the process remains elusive.
Simulations lend themselves conveniently to studying dissolution since they provide the spatio-temporal resolution that can be difficult to obtain experimentally.
Nevertheless, the complexity of various inter- and intra-molecular interactions require
careful treatment and long time scale simulations, both of which are typically hindered by computational expense.
Here, we use advances in machine learning potential methodology to resolve for the first time at an \textit{ab initio} level of theory the dissolution mechanism of NaCl in water.
The picture that emerges is that of a steady ion-wise unwrapping of the crystal preceding its rapid disintegration, reminiscent of crumbling. 
The onset of crumbling can be explained by a strong increase in the ratio of the surface to volume of the crystal.
Overall, dissolution is comprised of a series of highly dynamical microscopic sub-processes, resulting in an inherently stochastic mechanism.
These atomistic level insights now pave the way for a general understanding of dissolution mechanisms in other crystals, and the methodology is primed for more complex systems of recent interest such as water/salt interfaces under flow and salt crystals under confinement.
\end{abstract}

{\maketitle}

\section{Introduction}
\label{sec:introduction}
\begin{figure*}[ht]
    \centering
    \includegraphics[width=\textwidth]{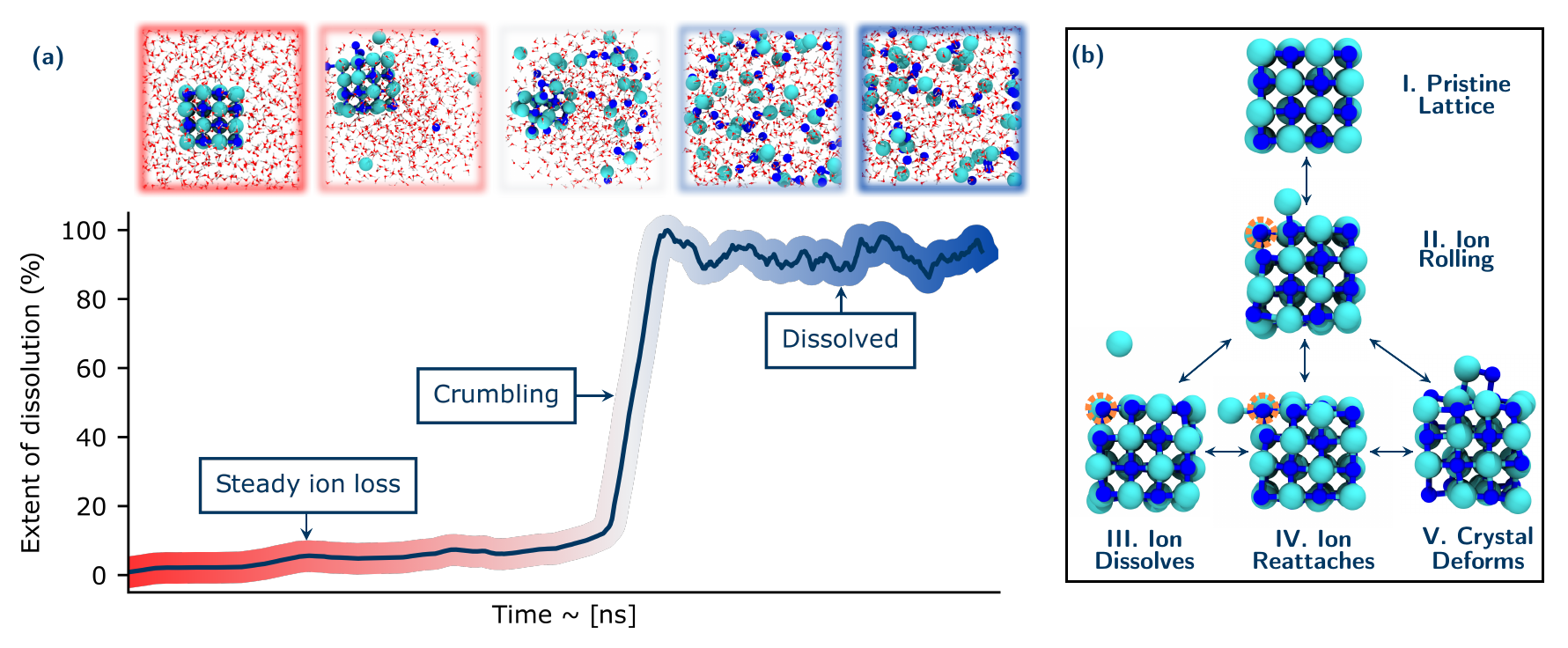}

    \caption{
	    \textbf{Overview of the NaCl dissolution process for a typical nanocrystal dissolution event.}
     Panel (a) shows the extent of dissolution of a NaCl nanocrystal in water, which involves a steady loss of ions followed by a rapid crumbling event. Relevant snapshots along the trajectory are shown above the plot. \ch{Na+} and \ch{Cl-} ions are coloured in blue and cyan respectively and oxygen and hydrogen atoms are coloured in red and white, respectively. (b) Sub-processes occurring during the dissolution. The vacancies left by the ions moving are shown by the orange circles. For clarity in panel (b) the water molecules present in the simulation cell are not shown. 
    }
    \label{fig:overview}
\end{figure*}

Understanding the dissolution of crystals is vital for a myriad of pressing modern day challenges, from technological issues such as battery science \cite{Suo2013/10.1038/ncomms2513} and water desalination \cite{Elimelech2011/10.1126/science.1200488} to drug bioavailability~\cite{Amidon1995/10.1023/A:1016212804288} and geochemical weathering \cite{Banfield1999/10.1073/PNAS.96.7.3404}.
The macroscopic process of dissolution is well described and understood from thermodynamics, where enthalpies of dissolution for example can be readily obtained from experimental techniques such as calorimetry \cite{Pitzer1884/10.1063/1.555709}.
However despite well understood general theories describing the conceptually reverse process of nucleation \cite{Sosso2016/10.1021/acs.chemrev.5b00744}, at a microscopic level our understanding of dissolution remains largely unresolved.

Dissolution is indeed an old problem \cite{Wulff1901/10.1524/ZKRI.1901.34.1.449},
however with recent experimental developments, it is an exciting time for the field.
Building on early atomic force microscopy studies of the water/NaCl interface by Xu et al. \cite{Xu1998/10.1016/S0039-6028(98)00199-X},
sophisticated imaging techniques such as scanning tunneling microscopy \cite{Peng2017/10.1088/1361-648X/aa537a} and Bragg coherent diffraction imaging \cite{Clark2015/10.1038/nmat4320} have given insight into low temperature surface ion dissolution events and the role of defects in the dissolution process. 
In addition, recent impressive high-resolution liquid cell transmission electron microscopy work has observed defect-mediated ripening of Cd-CdCl2 core-shell nanoparticles \cite{Zhang2022/10.1038/s41467-022-29847-8}.
The solid-liquid interface under flow has also been probed with a combination of surface-specific sum frequency generation spectroscopy and microfluidics experiments, revealing drastic changes to the equilibrium of dissolved ions.\cite{Lis2014/10.1126/SCIENCE.1253793}
Recent single-molecule atomic-resolution real-time electron microscopy experiments have captured in real time a NaCl nucleus emerging during nucleation \cite{Nakamuro2021/10.1021/jacs.0c12100} while atomic electron tomography has been used to monitor the nucleation of FePt nanoparticles over time \cite{Zhou2019/10.1038/s41586-019-1317-x}.
However in general, obtaining atomic scale dynamical information from experiment remains a highly challenging endeavour.

In contrast, computer simulations conveniently grant access to the \r{A}ngstrom and femtosecond scale resolution required, and have been applied to study the dissolution of a range of materials (see e.g. refs ~\onlinecite{%
    Silvestri2022/10.1021/acs.jctc.2c00787, %
    Aragones2012/10.1063/1.4728163, %
	Kurganskaya2013/10.1021/JP408845M,%
	Chen2014/10.1039/C4CP02375G,%
	Greiner2014/10.1021/mp500148q,%
	Ohlin2010/10.1038/nmat2585, %
	Leeuw1999/10.1103/PhysRevB.60.13792,%
	Dove2005/10.1073/pnas.0507777102,%
	Hofmann2012/10.1073/pnas.1208184109}).       %
Important insights have been obtained from these simulations that have significantly advanced the field,  such as the fact that 
dissolution rates are highly controlled by the crystal structure, and that the existence of site specific barriers to dissolution implies a non-constant dissolution rate throughout the process~\cite{%
	DeLaPierre2017/10.1002/anie.201701701, %
	Piana2005/10.1021/ja043395l, %
	Li2022/10.1038/s41467-022-28932-2,%
	Gao2013/10.1021/mp4000212,%
	Dove2007/doi.org/10.1063/1.2751917}.
Nevertheless, simulation studies up to now have had limitations.  
Notably, both force field (FF) and \textit{ab initio} studies have been limited to individual trajectories of dissolution events. 
However, a stochastic dissolution process is expected, as shown for example in larger scale studies in the geosciences \cite{Fischer2012/10.1016/j.gca.2012.09.011}.
Therefore it is important to establish if the sequence of dissolving atoms can be rationalised at an atomistic level from an ensemble of dissolution trajectories. 
In addition, nucleation pathways are strongly dependent on solution concentration \cite{Jiang2019a/10.1063/1.508428, Finney2022/10.1039/d1fd00089f,Lamas2021/10.1039/D1CP02093E}, raising  the question whether dissolution follows the same predictable mechanism or if it is highly sensitive to the conditions.

The questions noted above highlight the need for
computationally efficient explorations of multiple dissolution trajectories using a methodology that accurately describes the delicate changes in water-water and water-ion interactions that occur during dissolution.    
Fortunately, with the emergence of machine learning interatomic potentials (MLPs)
\cite{Kang2020/10.1021/acs.accounts.0c00472, Deringer2019/10.1002/adma.201902765,Behler2016/10.1063/1.4966192,Behler2021/10.1021/acs.chemrev.0c00868}, 
accurate and efficient potentials can be developed for the treatment of complex processes such as dissolution. 
Such potentials yield the accuracy of \textit{ab initio} methods but at a fraction of the computational cost.
Of the many flavours of MLP methodology, the approach reported in ref. ~\onlinecite{Schran2021/10.1073/pnas.2110077118}  is particularly suitable as it enables the automated development and validation of MLPs capable of accurately treating complex aqueous systems.

Given the opportunities presented by the recent developments in MLP technology,
we have performed a detailed study of NaCl dissolution. 
NaCl dissolution is a prototypical and widely studied system~\cite{%
        Lanaro2015/10.1021/jp512358s,
        Klimes2013/10.1063/1.4840675,%
        Liu2011/10.1039/C1CP21077G,%
        Holmberg2014/10.1039/c4cp00635f,
        Gaiduk2017/10.1021/acs.jpclett.7b00239, %
        Zhang2020/10.1038/s41467-020-16704-9, %
        Joswiak2018/10.1073/pnas.1713452115,
        Kobayashi2014/10.1063/1.4870417,%
        Bianco2022/10.1063/5.0083371,%
        Yang2005/10.1103/PhysRevE.72.012602,
        Zhang2022/10.1038/s41467-022-28538-8}, 
because of its ubiquity and significance to phenomena including biological intracellular reactions
	\cite{Flowers2015/10.1093/aob/mcu217}
and climate chemistry \cite{Simpson2007/10.5194/acp-7-4375-2007, Finlayson-Pitts2003/10.1021/cr020653t},
 not forgetting that two-thirds of the Earth is covered in salty water.
We find that the dissolution process is highly dynamic. Many stochastic sub-processes combine to give an overall crumbling mechanism, in which a steady period of ion-wise dissolution precedes the rapid concerted disintegration of the crystal.
This disintegration is governed by the steep increase of the surface to volume ratio of the crystal, corresponding to an unfavourable surface to bulk free energy -- borrowing from ideas of classical nucleation theory (CNT) -- that ultimately leads to the rapid collapse of the crystal.

\begin{figure*}[ht]
    \centering
    \includegraphics[width=\textwidth]{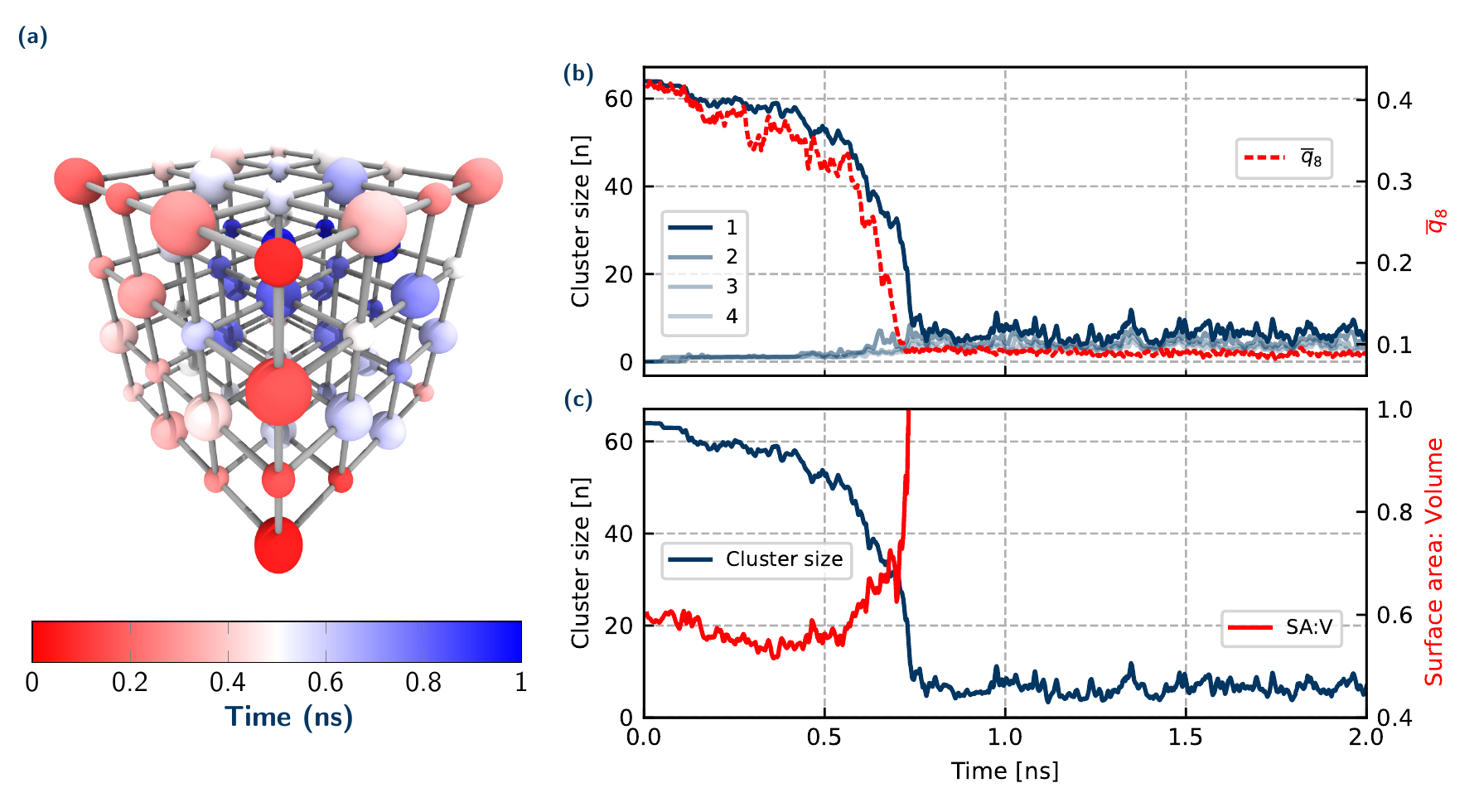}

    \caption{
	    \textbf{Summary of crystal and system properties for a dissolution trajectory at 2.42 mol/kg.}
	    (a) Time taken for individual ions to dissolve in a representative NaCl nano-crystal, where the larger ions are \ch{Cl-} and smaller ions are \ch{Na+}. Evolution over time of (b) cluster size of 4 largest ionic clusters and crystalline order parameter $\bar{{q}}_8$ (c) the ratio of surface area to volume (red, SA:V) of largest cluster (blue).
    }
    \label{fig:crumbling}
\end{figure*}

\section{The NaCl dissolution mechanism }\label{sec:overview}

Figure~\ref{fig:overview} provides an overview of a prototypical system considered and the NaCl dissolution process. 
Throughout this discussion, we show representative trajectories at one temperature but the phenomena observed hold for multiple trajectories at 330 K and 400 K.
Full details of the systems and simulation protocols are given in the Methods section. 
However, in brief, we considered 64 and 216 ion NaCl nanocrystals in simulation boxes containing 1250 to 625 water molecules (giving NaCl concentrations when dissolved of 1.42, 2.84 and 5.61\,mol/kg).
While we have not computed the solubility of the ML model, the experimental saturation of NaCl in water is 6.15\,mol/kg. All systems explored started out with a NaCl nanocrystal in pure water and the final dissolved state is likely undersaturated or at worst at a low supersaturation, thereby facilitating the complete dissolution of the crystal. 
A typical dissolution trajectory is shown in Figure~\ref{fig:overview} (a), with the extent of dissolution of the crystal monitored over time.
Relevant snapshots from the trajectory are shown in the upper panel.
After a steady period corresponding to ions dissolving from typically low coordinated sites of the crystal into solution, there is a rapid increase in the rate of dissolution.
This suggests the crystal reaches some critical point of instability after which it rapidly disintegrates or crumbles. 
Following this, the \ch{Cl-} and \ch{Na+} ions are fully solvated and the system has lost all crystalline order.
These general observations for the dissolution mechanisms are consistent over all conditions studied and reproducible over the ensemble of trajectories at a given condition.

The early stages of dissolution up to the onset of crumbling comprise a set of dynamic processes occurring in equilibrium as shown in Figure \ref{fig:overview} (b).
The dissolution of any ion from the crystal is first preceded by the ion rolling onto the surface of the crystal (Structures I. \& II.).
This process involves a simultaneous reduction in coordination of the rolling ion with its neighbours in the lattice until it is just coordinated with one counter ion, and a corresponding increase in coordination of the ion with water.
This behaviour is consistent with earlier AIMD studies on NaCl  and \ch{Li_xMn_2O_4}(001) surfaces.~\cite{ Liu2011/10.1039/C1CP21077G, Leung2017/10.1021/ACS.CHEMMATER.6B04429} 
Moreover, there is a difference between the rolling behaviour of \ch{Na+} and \ch{Cl-}: \ch{Na} remains much more stable on the surface, while \ch{Cl-} is much quicker to dissolve into solution.
This is in agreement with recent work by Silvastri et al. \cite{Silvestri2022/10.1021/acs.jctc.2c00787}, in which they show that \ch{Na+} ions are stable on terrace sites when they move from kink sites, while \ch{Cl-} has much weaker minima on terrace sites.
There are then several possibilities for the fate of this exposed ion:
It can become further solvated by water and completely dissolve from the crystal (Structure III.)
It can also roll back to its starting position and/or further roll to an orthogonal face (where it again has the possibility to dissolve from the crystal) (Structure IV.).
The dissolved ion can also rejoin the crystal (at not necessarily the same location as it originally emerged).
In these early stages before the rapid collapse of the crystal, other ions can also simultaneously roll and move about the crystal without necessarily any dissolution.
This results in a deformation of the crystal shown in Structure V.
The implications of these early dynamic processes with respect to the stochastic nature of dissolution will be discussed in detail in Section \ref{sec:stochastic}.

Figure \ref{fig:crumbling} (a) shows a schematic of the typical `unwrapping' of the crystal. 
The ions are coloured according to their time taken to dissolve.
Overall there is a preference for ions from low coordinated sites to dissolve first. 
We observe that dissolution is initiated at corner sites in the lattice, which has been previously noted in several papers~\cite{%
    Liu2011/10.1039/C1CP21077G,%
    Holmberg2014/10.1039/c4cp00635f,%
    Lanaro2015/10.1021/jp512358s,%
    Yang2005/10.1103/PhysRevE.72.012602,
    Hayakawaf1988/10.1351/pac198860081321}.
The dissolution of ions up to the point of crumbling proceeds in an approximately step-wise manner of ions of opposite charge (\ch{ Cl- }, \ch{ Na+ }, \ch{ Cl- }...), thereby minimising charge accumulation on the crystal lattice, confirming earlier force-field simulations on the initial stages of dissolution~\cite{%
    Yang2005/10.1103/PhysRevE.72.012602%
}.
At each dissolution step (depending on its location) for an ion to dissolve it must break 3 (corner), 4 (edge), 5 (face) or 6 (centre) ionic bonds.
Therefore the edge-wise unwrapping of the crystal shown in Figure \ref{fig:crumbling} (a) follows the hypothesis that at each step, the number of ionic bonds broken is minimised. 
This holds true up to the point of rapid disintegration, where all atoms irrespective of their current coordination dissolve within a few picoseconds.

These initial observations raise numerous questions regarding details of the crumbling mechanism proposed. Specifically: (i)
Does the crystal splinter into smaller crystalites or completely crumble into ions? 
(ii) Nucleation studies of NaCl in water, have shown that at very high concentrations there is an amorphous ionic structure before ions from solution form a nucleus \cite{%
Jiang2019a/10.1063/1.508428,%
Finney2022/10.1039/d1fd00089f%
}.
Therefore is there also evidence for an amorphous intermediate before the complete dissolution of the crystal?
(iii) At what point does the crystal disintegrate, and is this crumbling event determined by the size of the remaining crystalline cluster or something else?
To answer these questions, we now analyse the nature of the crumbling mechanism, and attempt to quantify and describe the driving force for this event.
Let us first address the question whether the remaining crystal splinters into smaller subclusters which is suggested in Ref. \citenum{Holmberg2014/10.1039/c4cp00635f}.
Another possibility is that it rapidly crumbles into individual ions. %
Monitoring the sizes of the 4 largest crystalline (ionic) clusters and the crystalline order of the largest cluster over time, as shown in Figure \ref{fig:crumbling} (b) gives insight into the exact mechanism of the crystal collapse.
Considering the sizes of the 4 largest ionic clusters, initially there is just one large cluster with 64 ions (i.e. the initial $4\times4\times4$ crystal), which gradually decreases in size corresponding to the steady period of ions dissolving.
At the point of crumbling, the number of ions in the largest cluster rapidly drops and immediately converges to a steady value of just one or two ions rapidly fluctuating over time (corresponding to short-lived ion pairs in solution).
Since at no stage is there a second cluster with some intermediate number of ions, this implies that there is a complete disintegration of the crystal, rather than something resembling a shattering process.

To understand if an amorphous transition precedes crumbling, the crystalline order of the ions in the system has been monitored over time.
$\bar{{q}}_8$ (defined in the SI) is a variation of the typical Steinhardt bond order parameter $q_8$, and averages the bond order vectors over the first shell of neighbouring ions to provide a measure of ordering of the ions in the system.
$\bar{{q}}_8$ (Figure \ref{fig:crumbling} (b)) qualitatively tracks the evolution of the largest cluster size, whereby initially there is a steady decrease (but still within the range expected for a crystal) corresponding to the crystal decreasing in size.
At a critical value of $\bar{{q}}_8$ $\approx$ 0.3, there is a sharp decrease in $\bar{{q}}_8$, corresponding to a rapid total loss of crystalline order in the system.
Therefore for the (low) concentration regimes we explore, our results are in agreement with the nucleation literature, where there is a single step order to disorder transition, with no amorphous intermediate.
We have already suggested that the onset of crumbling occurs when the crystal reaches a critical point of instability.
Across the simulated trajectories, there is a wide
spread of crystal sizes at which the disintegration occurs
(approximately 20-40 ions). 
So it is not size alone that governs the instability of the nanocrystal. 
To understand this behaviour  we considered the surface area to volume ratio of the convex hull of the largest cluster.
Panel (c) in Figure \ref{fig:crumbling} shows this surface area to volume ratio for the crystal over time up to the point of collapse.
This gives a measure of the extent of interaction of the ions in the crystal with water over time.
There is an initial decrease corresponding to a `rounding' of the crystal, to a relatively constant value up to the point of crumbling. (This rounding is even more pronounced for the larger $6\times6\times6$ crystal, Figure S10.)
At the onset of crumbling there is then a rapid increase in the surface area accessible to the water.
Our system is not in a concentration regime where reversible dissolution and nucleation is to be expected. 
However, with this ratio we can draw direct parallels with CNT. 
Within CNT, the free energy of formation of a crystal nucleus comprises a volume term representing the gain in free energy upon formation of a stable crystalline phase, and an unfavourable surface term arising from the cost of forming an interface.
The dominance of the surface area over the volume at the point of crumbling of the crystal lattice in this dissolution work is akin to a rapid increase in the (destabilising) surface free energy in CNT.
Therefore there is a delicate balance of intra- and inter-molecular interactions that ultimately determine the crystal stability.

We note that the cluster at the point of crumbling here is not equivalent to a critical nucleus in CNT. For the concentration regimes explored here, the critical nucleus would be expected to be much larger \cite{Jiang2019a/10.1063/1.508428}.
To further test the robustness of our proposed crumbling mechanism, we performed simulations on the same systems using the Joung Cheatham force field for ion-ion interactions with the SPC/E water model \cite{Benavides2016/10.1063/1.4943780}. This has a computed solubility of 3.7\, mol/kg, and therefore we can be sure that our systems are below saturation when fully dissolved for this model.
Full details and results are given in the SI, however we observe qualitatively the same crumbling mechanism, with similar cluster sizes at the point of crumbling. Therefore despite not having the solubility of the ML model, we can be confident that this mechanism is not simply a reverse of CNT.

In summary, this ratio is a simple parameter that explains the onset of disintegration of the crystal. It can also be used directly for other systems, thereby allowing for a simple and intuitive understanding of dissolution processes in general.

\section{Stochastic nature of the NaCl dissolution process}\label{sec:stochastic}

\begin{figure*}[ht]
    \centering
	\includegraphics[scale=0.9]{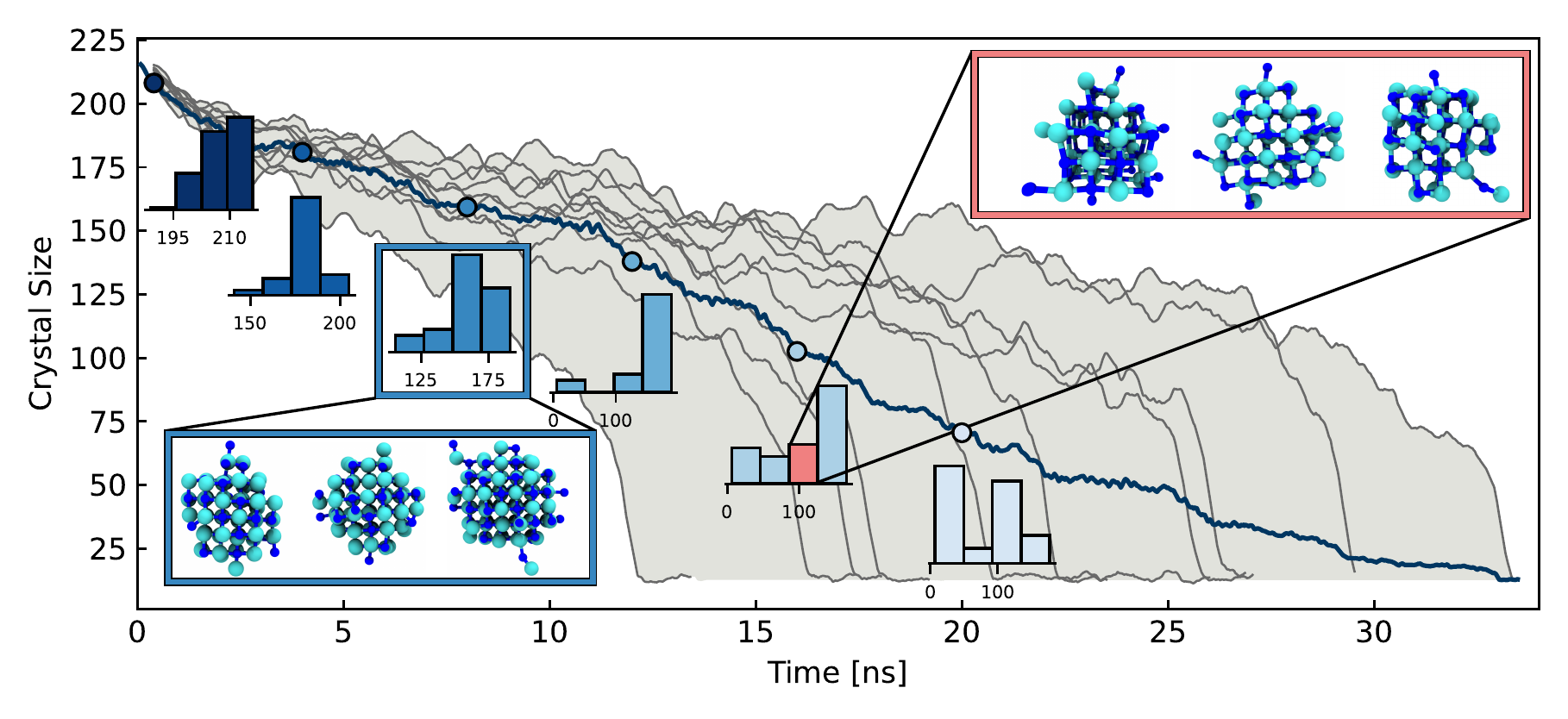}
    \caption{
	    \textbf{Overview of the stochastic nature of the NaCl dissolution process.}
     The main panel shows the crystal size over time for 10 simulations at 330 K initialised with random velocities for a $6\times6\times6$ nanocrystal. The range of fastest and slowest dissolving crystals is shaded in grey. The histograms show the distribution of crystal sizes across all trajectories for a given time window plotted along the mean crystal size (dark blue). Snapshots of crystals at 7 ns and with 87 ions are shown in the blue and red boxes respectively.}
    \label{fig:stochastic}
\end{figure*}

Thus far, we have described the overall crumbling mechanism  of NaCl dissolution in detail, which is completely general for multiple conditions of concentration and temperature.
However a pertinent question relevant for many fields is whether the dissolution mechanism is deterministic.
That is, given some initial conditions, can we predict at what crystal size/structure and after how long will the crystal dissolve?
Previous \textit{ab initio}  studies~\cite{%
Liu2011/10.1039/C1CP21077G,%
Holmberg2014/10.1039/c4cp00635f%
}
and larger scale FF-based studies on dissolution have limited their focus to individual trajectories, thus restricting the conclusions that can be drawn regarding the mechanism in this respect.
However in this work, access to multiple \textit{ab initio} quality trajectories via machine learning based simulations has enabled this issue to be addressed through a thorough statistical analysis over multiple trajectories.

Figure \ref{fig:stochastic} shows the time evolution of the crystal size distribution for 10 trajectories of a $6\times6\times6$ nanocrystal.   
The size distribution is initially very narrow at the beginning of the dissolution.
However as dissolution proceeds along the 10 trajectories, the distribution widens significantly.
After approximately 15 ns, there is a spread of over 100 ions in the crystal sizes along different trajectories.
While all trajectories follow the overall crumbling mechanism described in Section \ref{sec:overview}, this broadening of the distributions illustrates the diverse microscopic paths taken by the crystals \textit{en route} to dissolution.
We previously discussed in Section \ref{sec:overview} the highly dynamic equilibrium of microscopic subprocesses such as ions rolling on the crystal surface and lattice rearrangement occurring during the steady period of ion dissolution from the crystal.
These are rooted in inherent stochastic system fluctuations such as lattice and molecular vibrations (on the order of fs) and water dipole and hydrogen bond reorientations (on the order of ps).
Therefore the wide variation in the crystal structures at a given time (despite all simulations starting from the same pristine crystal structure) can be directly attributed to the stochastic nature of these microscopic sub-processes.
Given the relatively small sizes of the systems studied here, one would expect a much larger spread of times in macroscopic systems.
In addition to the range of crystal sizes for different trajectories at a given time, for a given crystal size, there are also numerous possible structures, as shown in the inset of Figure \ref{fig:stochastic}, with 3 representative examples of a crystal with 87 atoms.

These diverse structures again arise from a combination of the stochastic microscopic subprocesses.
The diversity in microscopic mechanism can also be observed for the other concentration conditions studied here. 
The largest $6\times6\times6$ crystal (5.61 mol/kg) takes much longer to dissolve and has a much broader range of dissolution times than the $4\times4\times4$ crystals (1.42 mol/kg and 2.84 mol/kg), as shown in Figure S10. 
This broad spread of times can be attributed to an increase in probability of ions rejoining the $6\times6\times6$ crystal, since it is surrounded by the lowest volume of water, but also there is a cumulative effect of each possible stochastic sub-process for every dissolution step.
While a distribution of dissolution times is to be expected, a 2 or 3-fold spread in times is particularly noteworthy.
In summary, the dissolution of NaCl therefore proceeds via an overall crumbling mechanism, within which, there is rich structural variety of the dissolving crystal owing to the inherently stochastic microscopic system fluctuations.

\section{Conclusion}
Despite dissolution being one of the most ubiquitous processes on earth, major gaps in understanding at the atomistic level persist. 
In this work we provide a rigorous description of the mechanism of NaCl dissolution, through the application of an automated framework to develop accurate MLPs.
Surpassing the extent of previous simulation work we have generated multiple trajectories ($>300$ ns in total) with \textit{ab initio} equivalent accuracy, under a range of conditions of NaCl dissolving.
We have established a general crumbling mechanism of NaCl dissolution, where following a steady period of ion-by-ion loss, the crystal reaches a critical point of instability and collapses in a concerted fashion.
The sequence of dissolving ions in the steady period of dissolution is such that the number of ionic bonds broken and net charge on the crystalline lattice is minimised.
The overall stability of a crystalline cluster in water can be straightforwardly reasoned by a delicate balance of interactions between the ions and water, measured via the surface area to volume ratio of the crystal.
This simple and physically intuitive concept applies to NaCl and to other ionic crystals in water and can readily be extended to non-ionic systems such as molecular crystals.
The crumbling mechanistic insights we have provided are primed for experimental measurement, including confinement controlled monitoring of the dissolution and Bragg coherant diffraction imaging, as have already been successfully applied to crystalline systems \cite{Lis2014/10.1126/SCIENCE.1253793,Nakamuro2021/10.1021/jacs.0c12100}. 

We have shown that the overall nature of the dissolution mechanism is highly stochastic, comprising a dynamic equilibrium of sub-processes such as ions rolling, dissolving, rejoining and deformations of the lattice in the crystal.
These arise from the stochastic hydrogen bond formation/breaking and thermal vibrations in the crystal and influence the relative stability of the crystal, and therefore the overall time taken for dissolution.
Given the sizes of the nanocrystals studied here, one could envisage a much larger spread of times in macroscopic systems such as those considered in the geosciences.
Indeed, studies of dissolution of crystalline and non-crystalline materials in the earth sciences have observed a broad variation in the rate constant of dissolution of up to 2-3 orders of magnitude.
\cite{
Fischer2012/10.1016/j.gca.2012.09.011,
Kurganskaya2021/10.1021/acsearthspacechem.1c00017}  %
which they attributed to the inhomogeneity of the initial crystal surface.
However our results suggest that this variance in rate is inherent to the system, regardless of initial condition. 

Such observations regarding the dynamic nature of the crystal are indeed very amenable to further experimental studies.
They suggest a large variation in dissolution mechanism under flow compared to confinement.
In the former case the ions are rapidly removed from the surface - similar to the lower ion concentration conditions we have shown here, while confinement essentially traps the ions, creating a high concentration of ions in the vicinity of the surface.
We propose that revisiting the microfluidics \cite{Lis2014/10.1126/SCIENCE.1253793} and electron microscopy \cite{Nakamuro2021/10.1021/jacs.0c12100} experiments described in the Introduction, with a focus on different concentration conditions would be highly insightful.

The accurate treatment of the electronic structure of the system required to capture relevant interactions, yet at a computational cost that facilitates the simulation of many dissolution trajectories is crucial for the observations made here.
The high transferability of the methodology and validation process is amenable to the study of other ionic salt systems.
Indeed, having now understood NaCl dissolution
-- in terms of a minimization of charge and of bonds broken, a balance of ion-solvent interactions determining crystal stability, and the dynamic nature of the dissolution process -- our work sets the stage for a generalised theory of dissolution of ionic crystals.
Computing the solubility of the ML model is also an important next step, given the large differences from experiment in solubilities predicted by different force field models \cite{Benavides2016/10.1063/1.4943780}.
Finally, this methodology and the insights we have now obtained are primed for application to  more challenging and complex systems of recent interest, including highly concentrated electrolyte solutions \cite{Groves2021/10.1021/acs.jpclett.0c03718,Smith2016/10.1021/acs.jpclett.6b00867} and electrolyte solutions under confinement \cite{Zhao2021/10.1038/s41467-021-25938-0}.
\section{Methods}

\textbf{Machine Learning Potential:} Machine learning potentials provide a direct functional relationship between atomic positions and forces/potential energy. 
This bypasses the otherwise computationally expensive requirement to solve Schr{\"o}dinger-like equations and systems of much larger length and timescales than feasible with AIMD based approaches are now accessible.
Building upon the seminal work of Behler and Parrinello in neural network based ML potentials \cite{Behler2007/10.1103/PhysRevLett.98.146401}, the recent development of an active learning approach using a committee of NNPs has enabled the systematic development and validation of NNP representations of the potential energy surface for complex aqueous systems \cite{Schran2020/10.1063/5.0016004, Schran2021/10.1073/pnas.2110077118}

Our model was iteratively trained on forces and energies from carefully chosen \textit{ab initio} molecular dynamics simulations in the generalised gradient approximation using the rev-PBE functional with Grimme's D3 dispersion correction \cite{Grimme2010/10.1063/1.3382344},
while using a Coulomb baseline to incorporate the correct long-range electrostatics, evaluated with particle mesh Ewald summation.
This baseline uses TIP3P model parameters \cite{Jorgensen1983/10.1063/1.445869} for water and point charges of +/- 1 for Na and Cl. 
This setup has been shown to be accurate for aqueous systems, and further details on functional selection and long-range electrostatics are described in the SI.

The model was iteratively improved over multiple generations such that the relevant configuration space was accurately covered in the overall training set with an energy and force root mean square error (RMSE) of 1.3 meV per atom and 38.0 meV/\AA{}, respectively. 
The model was extensively validated in a series of tests incorporating static and dynamical  properties of selected test systems of which full details are also given in the SI.

\textbf{MD simulations:} 
The systems used in all dissolution simulations described comprised a $4\times4\times4$ ($6\times6\times6$) NaCl nanocrystal with 32 (108) Na/Cl atoms using a computed lattice constant of 5.72 \AA{} \cite{Liu2009/10.1063/1.3152845}.
The $4\times4\times4$ nanocrystal was surrounded by 625 (1250) water molecules in a cubic simulation cell of side length 27.8 \AA\: (34.1 \AA), giving dissolved NaCl concentrations in water of 2.84 mol/kg and 1.43 mol/kg respectively. The $6\times6\times6$ crystal was surrounded by 1068 waters in a cubic simulation cell of side length 34.1 \AA \:  giving a dissolved NaCl concentration in water of 5.61 mol/kg.
MLP simulations were all carried out using the CP2K/Quickstep code in the NVT ensemble and at a constant temperature of 330 K (chosen to be consistent with previous literature \cite{Liu2008/10.1021/ja8014296}) maintained using the CSVR thermostat \cite{Bussi2007/10.1063/1.2408420}.
At each concentration, 10 trajectories were simulated using the same initial configuration and initialised with random velocities drawn from the Boltzmann distribution at the target temperature of 330 K. 
Simulations were run for over 15 ns (5.61 mol/kg), 2 ns (2.84 mol/kg) and 1 ns (1.42 mol/kg) until the crystal was fully dissolved.
The same simulation procedure was carried out for 10 trajectories of the 2.42 mol/kg concentration at 400 K for further validation of the generalisability of our conclusions. These results are included in the SI.
Overall over 300 ns of \textit{ab initio} quality machine learning based simulations were performed, far beyond the capabilities of AIMD simulations.
Further details for computing parameters including cluster size, $\bar{{q}}_8$ and surface area to volume ratio of the crystal are given in the SI.

\section{Acknowledgments}
We are grateful to Daan Frenkel for helpful discussions. We thank Julian Gale for his insightful comments on the manuscript. N.O.N acknowledges financial support from the Gates Cambridge Trust. C.S acknowledges financial support from the Alexander von Humboldt Stiftung. S.J.C is a Royal Society University Research Fellow (URF\textbackslash R 1\textbackslash 211144) at the University of Cambridge. We are grateful for computational support and resources from the UK Materials and Molecular Modeling Hub which is partially funded by EPSRC (Grant Nos. EP/P020194/1 and EP/T022213/1). We are also grateful for computational support and resources from the UK national high-performance computing service, Advanced Research Computing High End Resource (ARCHER2). Access for both the UK Materials and Molecular Modeling Hub and ARCHER2 were obtained via the UK Car-Parrinello consortium, funded by EPSRC grant reference EP/P022561/1.

%
%
%

\section*{References}

%

\end{document}


\def\mytitle{%
Crumbling Crystals: On the Dissolution Mechanism of NaCl in Water
}
\title{Supporting Information for: \mytitle}

\author{Niamh O'Neill}%
\affiliation{%
Yusuf Hamied Department of Chemistry, University of Cambridge, Lensfield Road, Cambridge, CB2 1EW, UK
}

\author{Christoph Schran}%
\email{cs2121@cam.ac.uk}
\affiliation{%
Yusuf Hamied Department of Chemistry, University of Cambridge, Lensfield Road, Cambridge, CB2 1EW, UK
}
\affiliation{%
Thomas Young Centre, London Centre for Nanotechnology, and Department of Physics and Astronomy, University College London, Gower Street, London, WC1E 6BT, United Kingdom
}%

\author{Stephen J. Cox}%
\affiliation{%
Yusuf Hamied Department of Chemistry, University of Cambridge, Lensfield Road, Cambridge, CB2 1EW, UK
}%

\author{Angelos Michaelides}%
\email{am452@cam.ac.uk}
\affiliation{%
Yusuf Hamied Department of Chemistry, University of Cambridge, Lensfield Road, Cambridge, CB2 1EW, UK
}
\affiliation{%
Thomas Young Centre, London Centre for Nanotechnology, and Department of Physics and Astronomy, University College London, Gower Street, London, WC1E 6BT, United Kingdom
}%

{\maketitle}
{\tableofcontents}

\onecolumngrid

\section{Selecting DFT functional}
Two modern DFT functionals that have been shown to perform well for liquid water simulations \cite{Zheng2018/10.1063/1.5023611,Galib2017/10.1063/1.4986284} - the dispersion corrected GGA functional revPBE-D3 \cite{Perdew1996/10.1103/PhysRevLett.77.3865,Zhang1998/10.1103/PhysRevLett.80.890} and the strongly constrained and appropriately normed (SCAN) meta-GGA functional \cite{Sun2015/10.1103/PhysRevLett.115.036402}- were selected for an initial benchmark study to determine the most suitable XC functional for the \textit{ab initio} simulations upon which the MLP would be trained.
It should be noted that dispersion corrected hybrid functionals  have been shown to provide slightly better results for liquid water simulations \cite{Marsalek2017/10.1021/acs.jpclett.7b00391,Gaiduk2018/10.1021/acs.jpclett.8b01017}; however initial simulations with revPBE0-D3 showed there is approximately a six-fold increase in computational cost, rendering longer time-scale AIMD simulations unfeasible for the envisaged system sizes.
\newline

\textbf{Na/Cl ion in water}

\begin{figure*}[ht]
    \centering
    \includegraphics{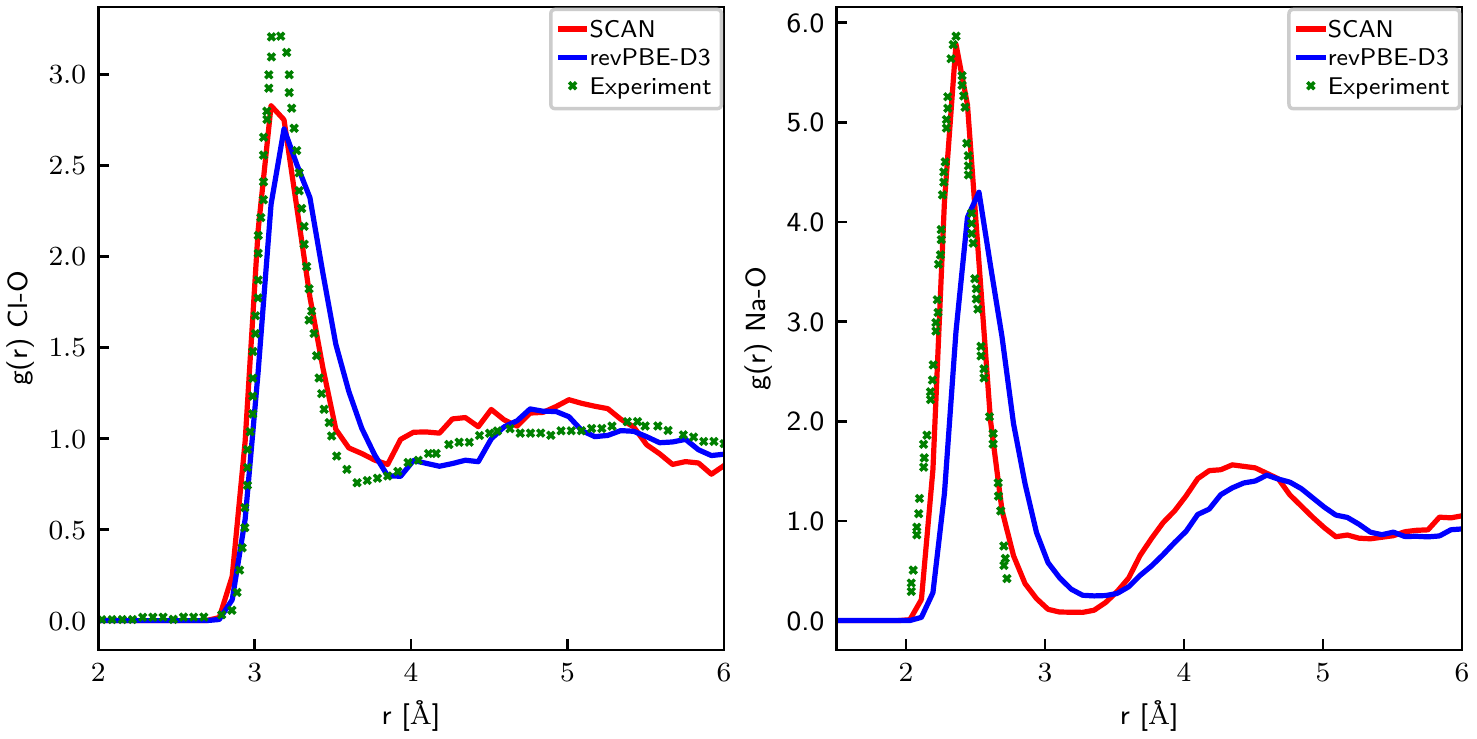}

    \caption{Radial distribution function g(r) for Cl-O (left) and Na-O (right) comparing revPBE-D3 and SCAN with experiment. Cl-O experimental RDF obtained from neutron diffraction data \cite{Soper2006/10.1016/j.bpc.2006.04.009} and Na-O experimental RDF obtained from X-Ray diffraction data \cite{Duignan2020/10.1039/c9cp06161d}.
    }
    \label{fig:rdf-ions}
\end{figure*}

To study the performance of SCAN and revPBE-D3 for ion-water simulations, four separate AIMD simulations were performed for both revPBE-D3 and SCAN for \ch{Na+} in water and \ch{Cl-} in water. 
The Na(Cl)-O RDF g(r) are compared to experimental values in Figure \ref{fig:rdf-ions} and Table \ref{tab:rdf}. %
Both SCAN and revPBE-D3 are in good agreement with the position of the first experimental Cl-O peak ($\mathrm{R^{1}_{max}}$), where both are within 0.07 \AA{} of the experimental data.
However the height g($\mathrm{R^{1}_{max}}$) is underestimated by both by approximately 15\%.
SCAN and revPBE-D3 both qualitatively agree with the position of the second peak maximum, but they both tend to overstructure compared to experiment.
The experimental Na-O peak is more accurately reproduced by SCAN than revPBE-D3.
The revPBE-D3 peak is shifted 0.17 \AA{} from the experimental value, and the peak height is 27\% lower.
This is also in agreement with previous studies, where SCAN has been shown to outperform GGA functionals in reproducing the sodium-water solvation structure \cite{Duignan2020/10.1039/c9cp06161d}.
Despite a lack of experimental data for the second solvation shell, revPBE-D3 is in much closer qualitative agreement with SCAN than for the first peak.
\begin{table}[h]
\caption{Summary of structural properties of solvated sodium/chloride ions obtained from RDFs g(r) computed from AIMD simulations with SCAN and revPBE-D3 at 300 K. Experimental reference data for Cl-O and Na-O were obtained from neutron diffraction \cite{Soper2006/10.1016/j.bpc.2006.04.009} and X-Ray diffraction \cite{Galib2017/10.1063/1.4986284} data respectively. Difference from experiment is shown in brackets.}
\centering
\begin{tabular}{@{}llllllll@{}}
\toprule
 & \multicolumn{3}{c}{Cl-O} &  & \multicolumn{3}{c}{Na-O} \\ \midrule
 & SCAN & revPBE-D3 & Experiment &  & SCAN & revPBE-D3 & Experiment \\ \cmidrule(r){1-4} \cmidrule(l){6-8}
$\mathrm{R^{1}_{max}}$ (\AA{}) & \multicolumn{1}{c}{3.10 (0.07)} & \multicolumn{1}{c}{3.19 (0.02)} & \multicolumn{1}{c}{3.17} & & \multicolumn{1}{c}{2.36 (0.00)} & \multicolumn{1}{c}{2.53 (0.17)} & \multicolumn{1}{c}{2.36}  \\
g($\mathrm{R^{1}_{max}}$) & \multicolumn{1}{c}{2.83 (0.62)} & \multicolumn{1}{c}{2.70 (0.51)} & \multicolumn{1}{c}{3.21} & & \multicolumn{1}{c}{5.79 (0.08)} & \multicolumn{1}{c}{4.30 (0.43)} & \multicolumn{1}{c}{5.87}  \\
 \bottomrule
\end{tabular}
\label{tab:rdf}
\end{table}

\begin{figure*}[ht]
    \centering
    \includegraphics{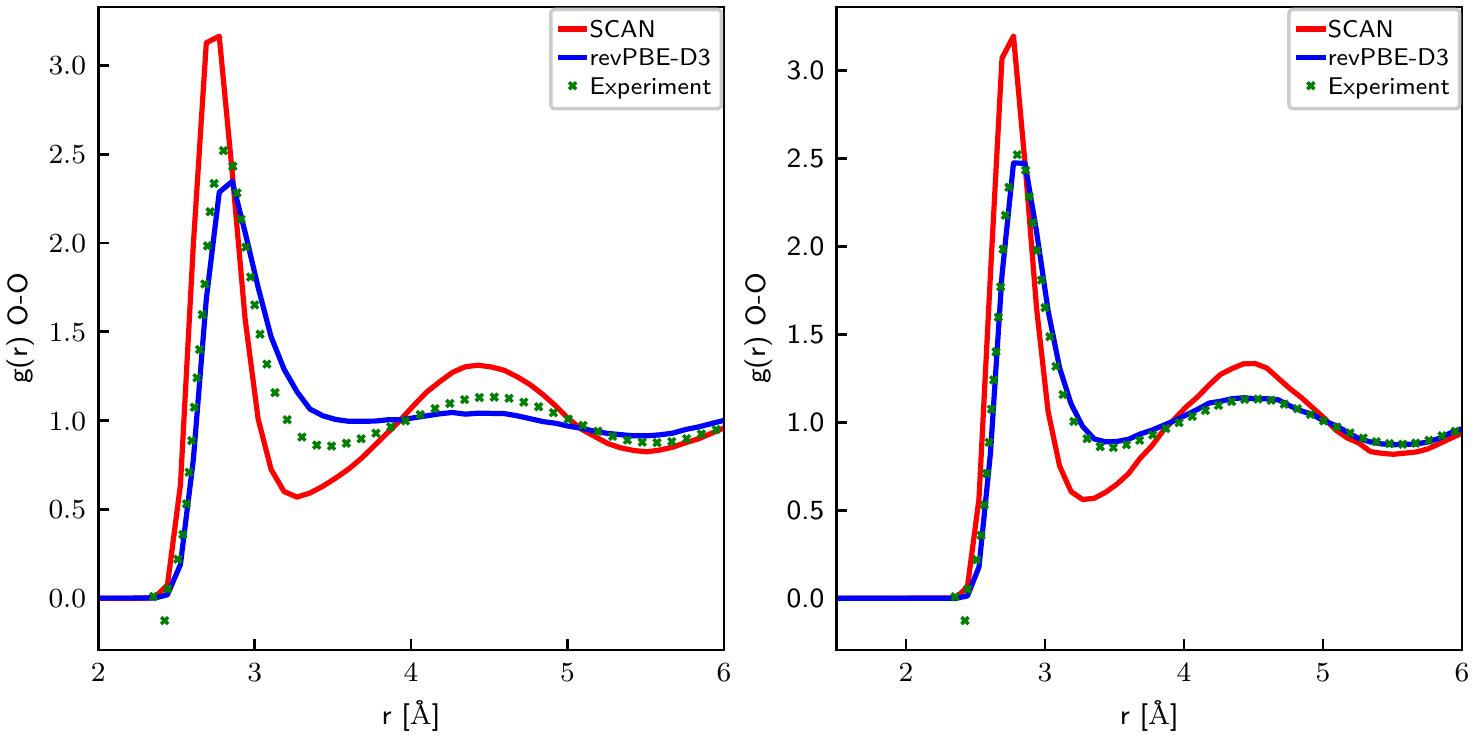}
        \caption{Radial distribution function g(r) for O-O for \ch{Cl-}/\ch{H2O} (left) and \ch{Na+}/\ch{H2O} (right). Compared to experimental data taken from Ref.~\citenum{Skinner2013/10.1063/1.4790861}.}
    \label{fig:rdf-water}
\end{figure*}

The O-O RDF for both solvated sodium and chloride ion systems was also computed and
compared to experiment in order to evaluate the performance of the two functionals for water.
This analysis is shown below in Figure \ref{fig:rdf-water}. In both cases, SCAN significantly overstructures water compared to revPBE-D3 and experiment, with a first peak maximum approximately
25\% greater than experiment. This is consistant with literature results, in which Duignan et al. compared the performance of revPBE-D3 and SCAN in simulations of liquid water at 300 K \cite{Duignan2020/10.1039/c9cp06161d}. revPBE-D3 reproduced well the experimental O-O radial distribution function (RDF), with the first peak maximium and magnitude coinciding to within 0.1 \AA{} and 2\% respectively, while SCAN overstructured water, with a first peak RDF maximum approximately 20\% larger than experiment. Galib et al. also report a computed water density with revPBE-D3 of 0.962 ± 0.029 $\mathrm{g cm^{-3}}$ within 4\% of the experimental density of 0.997 $\mathrm{g cm^{-3}}$ \cite{Galib2017/10.1063/1.4986284}, compared to a computed density by Wiktor et al. of 1.05 $\mathrm{g cm^{-3}}$ \cite{Wiktor2017/10.1063/1.5006146}.

In summary, the decision between SCAN and revPBE-D3 is not completely clear-cut and both functionals have advantages and disadvantages. 
revPBE-D3 performs better for liquid water than SCAN. 
With respect to the solvated ions, revPBE-D3 is on par with SCAN for the Cl-O interactions, however performs significantly poorer for Na-O.
Nevertheless, revPBE-D3 reproduces experimental data much better than SCAN with respect to the O-O RDF.
As a footnote, SCAN is also two-thirds more expensive than revPBE-D3.
Therefore, the AIMD trajectories generated using revPBE-D3 were chosen for the initial development of the C-NNP training set and revPBE-D3 was used for generation of subsequent AIMD data and during the active learning stages of the MLP development.

\section{Development of Machine Learning Potential}\label{sec:mlp-devel}

\subsection{Automated Work Flow}
\begin{figure*}[h!]
    \centering
	\includegraphics[width=\textwidth]{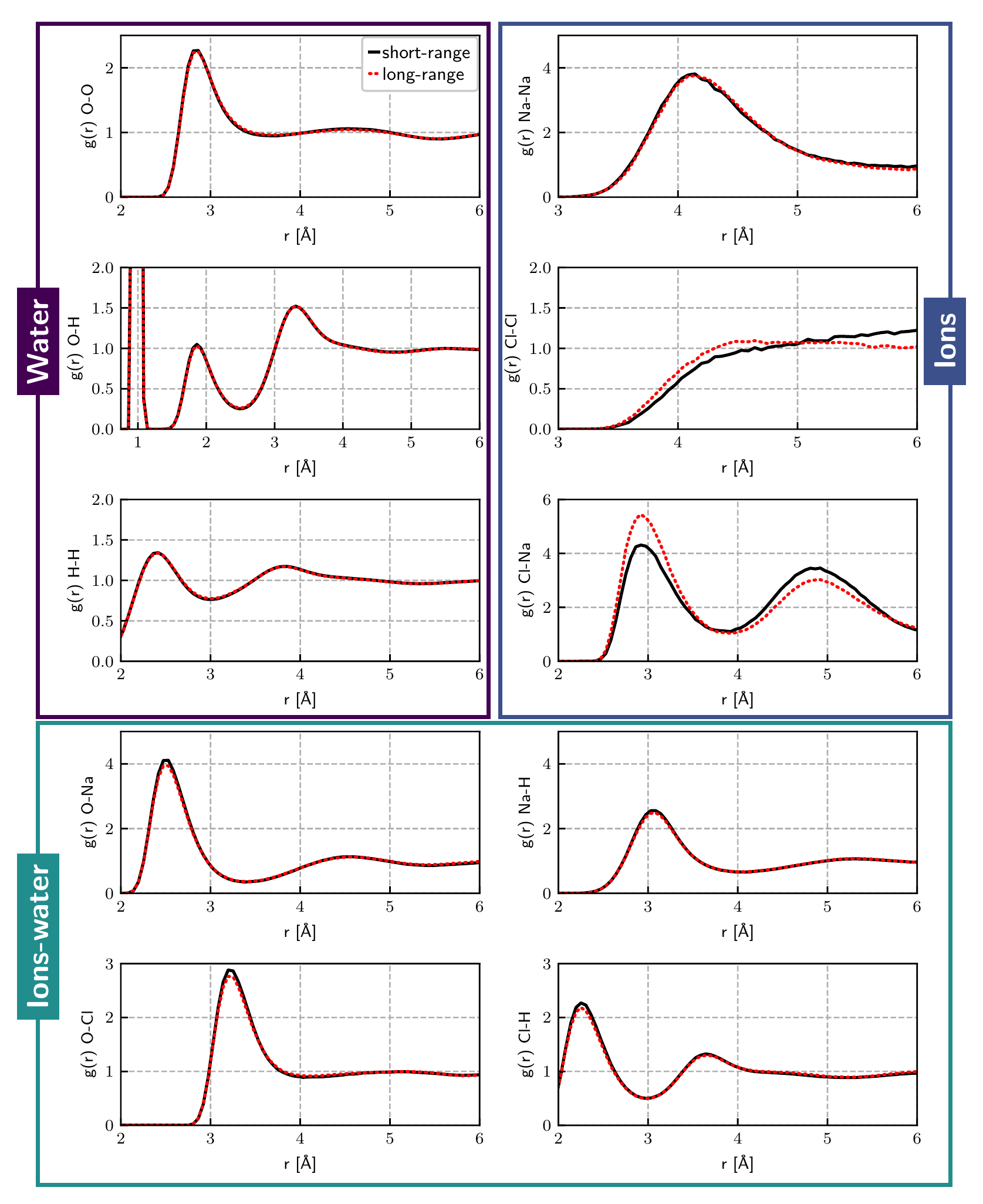}

    \caption{Summary of iterative development of MLP over 5 generations. The number of configurations included from each AIMD simulation is given in Generation-01. Thereafter, the number of configurations is the total number of structures in the training set for the given generation. 
    }
    \label{fig:generations}
\end{figure*}

The procedure for developing the committee neural network potential (C-NNP) was followed as described in Ref~\citenum{
	Schran2021/10.1073/pnas.2110077118}.
The general approach taken in developing the C-NNP was to treat different relevant systems and conditions individually with active learning, generating a training set for each condition and then combining the individual training sets to generate the final model.
The model was iteratively improved over five generations of such active learning rounds, with each generation targeting a specific region of phase space.
The overall development over these five generations of the model is summarised in Figure \ref{fig:generations}.
An initial potential was trained on forces and energies obtained from \textit{ab initio} simulations described in Section \ref{sec:simulations} comprising bulk water, individual \ch{Na+}/ \ch{Cl-} ions in water, an \ch{Na+}/ \ch{Cl-} ion pair in water and a small $2\times2\time2$ nanocrystal in water.
This model was then used under different temperatures to generate structures along the dissolution trajectory of the prototypical 4x4x4 nanocrystal (Generation 2).
Later generations of the model targeted solution and intermediate structures along the dissolution trajectory (Generations 3+4) and highly concentrated solutions (Generation 5).
The development of the model used the Python AML active learning package described in Ref.~\citenum{
	Schran2020/10.1063/5.0016004}.
For a given active learning iteration, 20 random structures from a reference trajectory were used to initialise the model. %
After training 8 NNP members, forces and energies of 2000 randomly selected structures from the reference trajectory were predicted to ascertain the force and energy committee disagreements. %
20 structures with the largest mean force disagreement were added to the training set for the next round of active learning.
Convergence was reached when new structures added to the training set did not improve the committee disagreement between points already in the training set, indicating the training set was sufficiently diverse.
The training set of the final C-NNP model contained 1608 structures. 
The overall force and energy training root mean square error (RMSE) for the model are 38.0 meV/\AA{} and 1.3 meV/atom respectively.
Given the complexity of the system under study, this compares well to a RMSE of 70 meV/\AA{} quoted in Ref.~\citenum{Morawietz2016/10.1073/pnas.1602375113}
for a simpler system of large scale MD simulations of water using NNPs trained on DFT.

\subsection{Details of Model}
The chemical environment around each atom was described using a general set of atom-centered symmetry functions~\cite{Behler2011/10.1063/1.3553717}.
There are 10 radial and 4 angular functions for each pair and triple of atoms, following Ref.~\citenum{Schran2021/10.1073/pnas.2110077118}.
All symmetry functions used a cutoff function of angular cosine form with a cutoff radius of 12 Bohr.
The committee was comprised of 8 NNP members, of identical architecure with 2 hidden layers and 25 neurons in each layer.
In all cases, random sub-sampling was performed to introduce variability between the committee members, where 10\% of the total set of structures were discarded.
The weights and biases of the NNPs were optimised using the n2p2 code \cite{Singraber2019/10.1021/acs.jctc.8b01092}.
Individual models during active learning were optimised for 15 epochs, while the final C-NNP model used in simulations was optimised for 50 epochs.
To ensure that long-range effects are accurately captured by the machine learning potential, we explicitly incorporated long-range effects beyond the cutoff of the symmetry functions (12 Bohr).
The predicted energy can in general be written as a sum of short range and long range contributions ($E_{sr}$ and $E_{coul}$ respectively):	$E_{\mathrm{tot}} = E_{\mathrm{sr}}+E_{\mathrm{coul}}$.
The long-range model was thus trained on the difference between the standard short-ranged model and the Coulomb contribution, calculated using point charges of +/- 1 respectively for Na and Cl and using TIP3P model parameters for water \cite{Jorgensen1983/10.1063/1.445869}.
We used this model in all production simulations, where the  Coulomb contributions were explicitly included via particle mesh ewald summation.
Details on the validation of the final model and the role of long range interactions in these simulations are presented in the next Section.

\FloatBarrier

\subsection{Validation}
The system studied is complex, so care was taken to set up a series of validation tests that incorporates relevant subsystems and properties.
\subsubsection{$2 \times 2 \times 2$ NaCl nanocrystal}
\begin{figure*}[ht]
    \centering
    \includegraphics{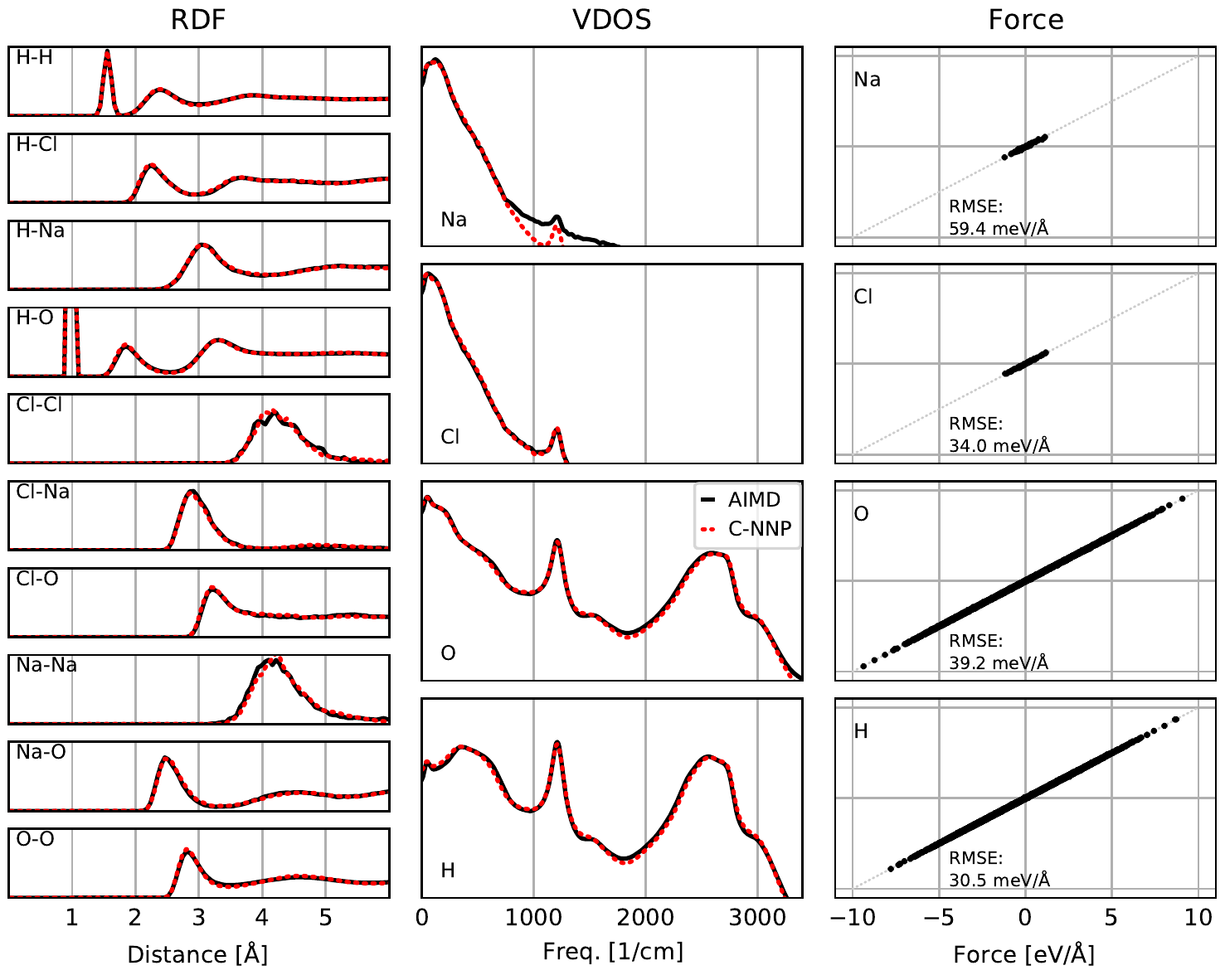}

    \caption{Benchmark tests of long-range C-NNP model predictions of structural and dynamical properties of a $2 \times 2 \times 2$ NaCl nanocrystal in water as well as force prediction. The comparison between AIMD and the C-NNP RDFs, VDOS and force correlation are shown in the left, middle and right panels respectively. 
    }
    \label{fig:2x2x2-lr}
\end{figure*}

The static and dynamical properties of the largest accessible NaCl nanocrystal in water for AIMD simulations ($2\times2\times2$ NaCl surrounded by 229 waters - 695 atoms) was analysed using a benchmarking scheme proposed in Ref.~\citenum{Schran2021/10.1073/pnas.2110077118}, comparing AIMD to the C-NNP predictions and shown in Figure \ref{fig:2x2x2-lr}.
The radial distribution functions (RDFs) $g(r)$ for all the atom species of the system gives a good assessment of the ability of the C-NNP to predict thermodynamic properties.
The vibrational density of states (VDOS) for each element gives information on the vibrational modes of the system, and thus the dynamics of inter- and intra-atomic interactions.
The forces on the atoms are sensitive to the local environment, and give additional insight via which the quality of the ML model can be evaluated.
RDF and VDOS predictions are all in essentially perfect agreement with the AIMD reference trajectory while the force predictions and the DFT reference are very well correlated.
It should be noted that the slight discrepancy in the VDOS between the C-NNP and AIMD at higher frequencies is likely due to the large plane wave cutoff required in the DFT calculations to obtain converged Na forces (Discussed in further detail in Section \ref{sec:simulations}).
Nevertheless the symmetry between Na and Cl VDOS spectra is reassuring.
\FloatBarrier

\subsubsection{Validation of Dissolution Process}
\begin{figure}[ht]
    \centering
    \includegraphics{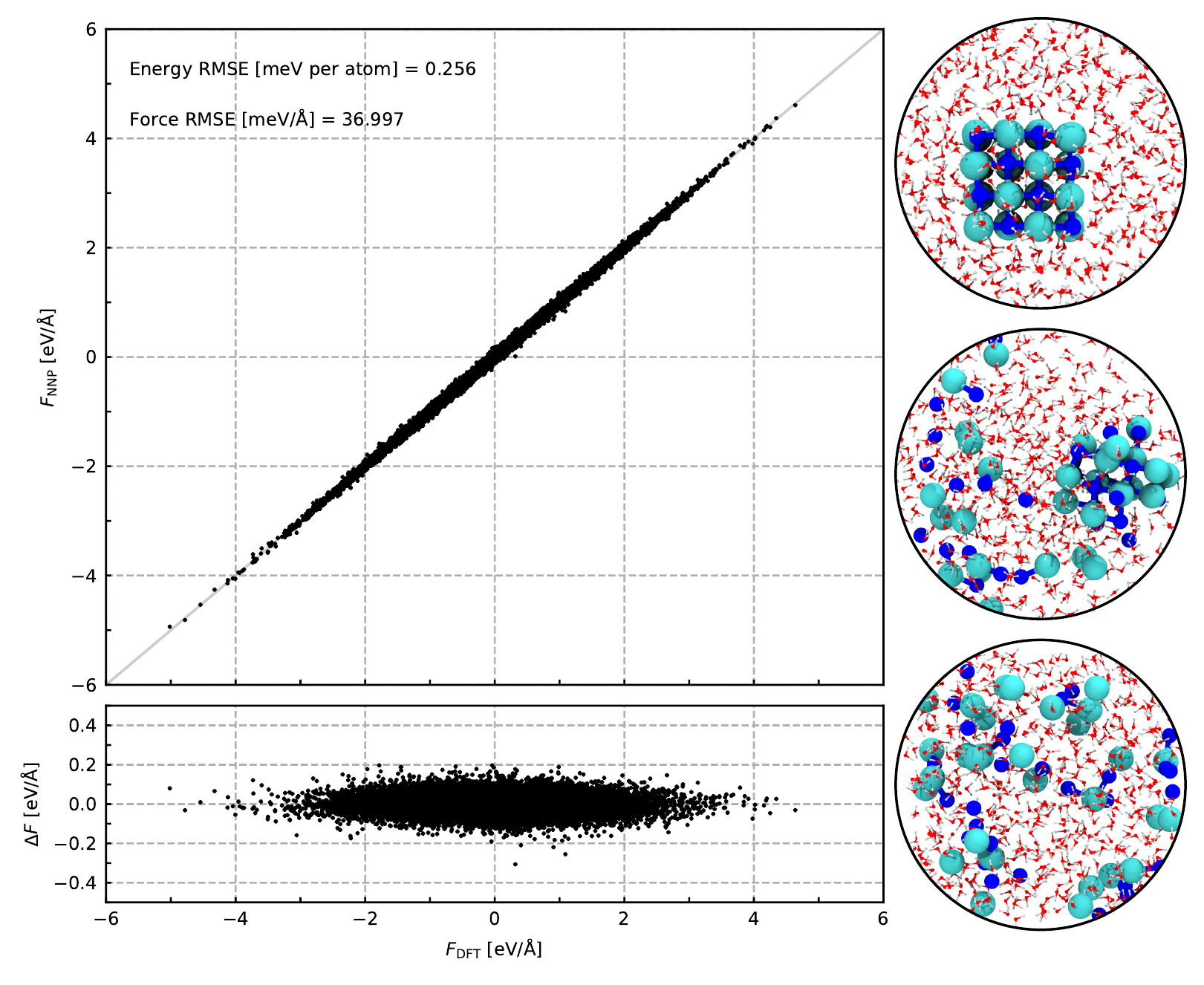}
    \caption[Correlation plot for long-range C-NNP model predicted forces and corresponding reference DFT forces.]{Correlation plot for C-NNP predicted forces and corresponding reference DFT forces, with light grey line showing a perfect correlation coefficient of 1. Representative snapshots of the variety of structures tested are shown on the right hand side of the figure.}
	\label{fig:rmse}
\end{figure}

As the most rigorous of tests for the C-NNP we validate the ML model for the actual dissolution process of a 4x4x4 nanocrystal in solution by comparing the force and energy predictions of the model and the corresponding DFT forces and energies for a set of structures uniformly selected along the dissolution process.
It is particularly valuable since the model has not been explicitly trained on these structures.
First, an extensive MD simulation with the developed model was performed in order to sample the dissolution process of the 4x4x4 nanocrystal in solution.
A variety of structures were selected to capture all of the relevant configurations along the trajectory; from the intact NaCl crystal lattice, to the initial stages of dissolution to the end point of the fully solvated ions in solution.
Next, the DFT reference energies and forces of these 50 structures were computed.
The resulting correlation of the forces is shown in Figure \ref{fig:rmse}.
The RMSE values for both forces (37.0 meV/\AA{}) and energies (0.3 meV/atom) compare very favourably to similar reactive systems, which have been studied using machine learning potentials, such as the work by Behler et al. in Ref.~\citenum{Quaranta2017/10.1021/ACS.JPCLETT.7B00358} who quote a force and energy RMSE for a model describing proton transport at ZnO/\ch{H2O} interfaces of 140.4 meV/\AA{} and 1.0 meV/atom respectively.

\subsection{Long-range Effects}
\begin{figure}[ht]
    \centering
    \includegraphics{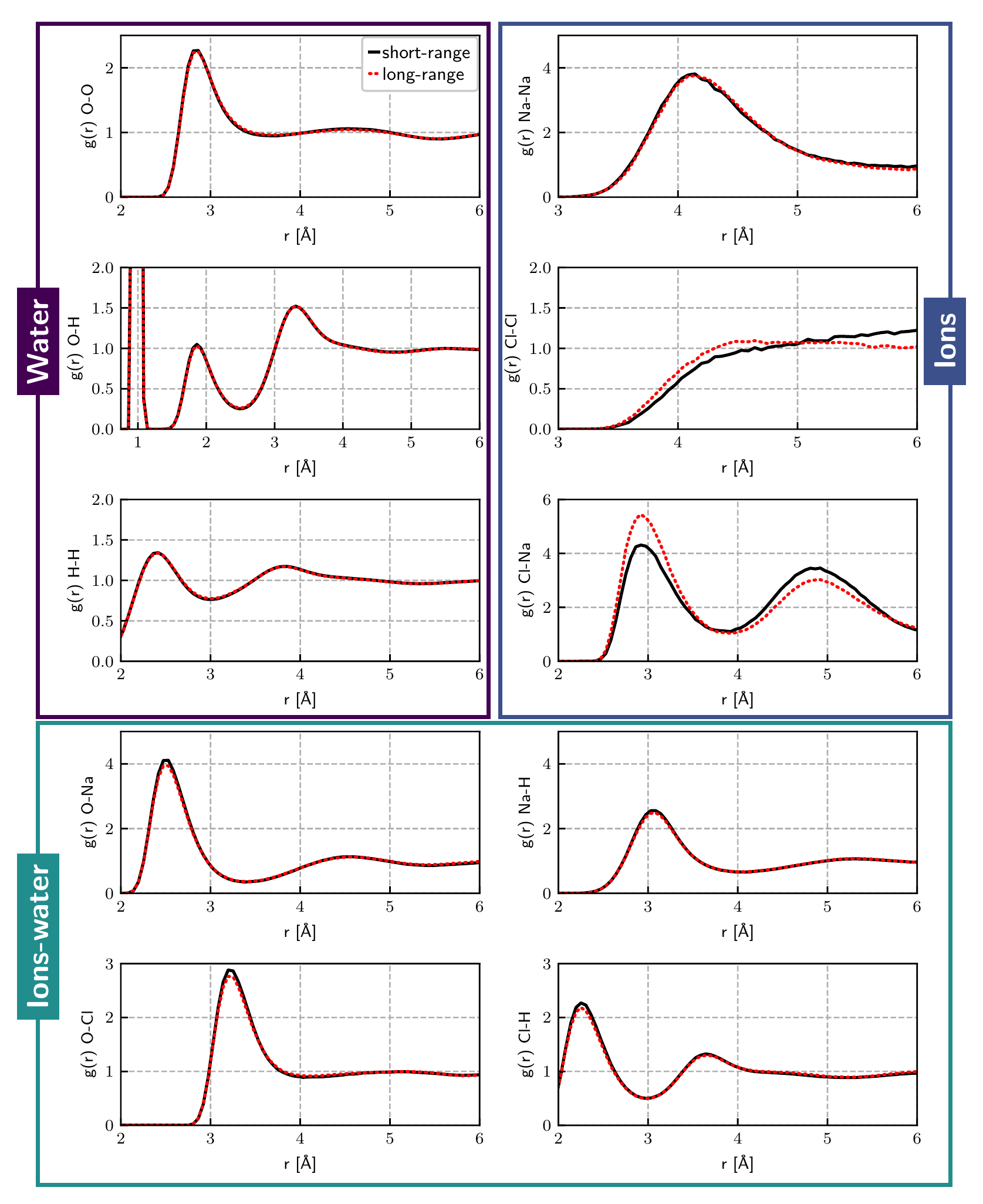}
    \caption[Comparison of long and short-range models in predicting NaCl solution properties.]{Comparison of long and short-range models in predicting NaCl solution properties. RDFs for 2.48 mol/kg NaCl solution for both long- and short-range model predictions are shown in red and black respectively.}
        \label{fig:long-range}
\end{figure}

In order to evaluate the effect of explicitly incorporating long range effects into the model structural properties (RDFs) of the 2.48 mol/kg NaCl solution were computed. These are compared to the short range model in Figure \ref{fig:long-range}. Both predictions are in very good agreement for all cases of water, ions and ion-water RDFs. Nevertheless small discrepancies between long- and short-range model predictions in the cases of the Cl-Cl and Na-Cl RDFs prompt us to use the long-range model for production runs, given the additional physical information included in the model. In general however this analysis suggests that overall for our simulated systems, long-range effects do not have a substantial impact. 

\FloatBarrier
In summary this set of validation tests confirm the overall quality and accuracy of the C-NNP model.
The model has performed comparably to or exceeded literature values on a variety of systems and properties that are very relevant for the dissolution process, from small scale studies of single ions in solution to systems sizes and interactions that will be explored in the detailed analysis of the dissolution process.

\section{Molecular dynamics simulations}\label{sec:simulations}
\subsection{System setups}
All simulations were carried out in an orthorhombic simulation cell with periodic boundary conditions in $x,y,z$ directions.
Details of the simulation setups for all systems are shown in Table \ref{tab:setups}.

\begin{spacing}{.8}
\begin{longtable}{lll}
%
%
\caption{Details of systems simulated. L is (unless stated) the cubic box dimensions in \AA{}.} \label{tab:systems}\\
%
\multicolumn{1}{l}{\textbf{System}} & \multicolumn{1}{l}{\textbf{Simulation Details}}  & \multicolumn{1}{l}{\textbf{Snapshot}}\\ \midrule
	Bulk H$_2$O &
	\begin{tabular}[c]{@{}l@{}}N$\mathrm{_{water}}$ = 64\\ L = 12.42\,\AA{} \end{tabular}&   
		\begin{minipage}{.3\textwidth}
			\includegraphics[trim={3cm 2cm 1.7cm 2cm},clip,height=23mm]{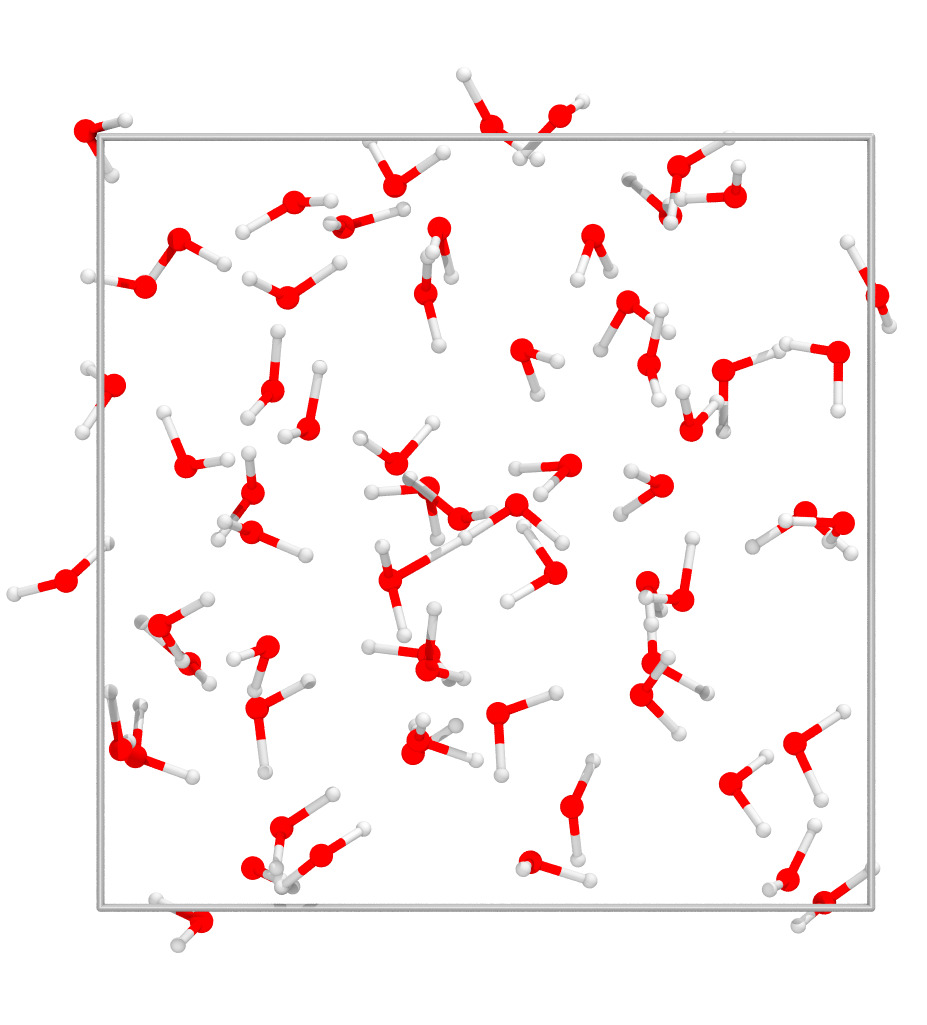}
    		\end{minipage} \\ \midrule
	\ch{Na+} / H$_2$O & 
	\begin{tabular}[c]{@{}l@{}}N$\mathrm{_{water}}$ = 63\\ N$_{\mathrm{Na}}$ = 1\\ L = 12.42\,\AA{}\end{tabular} &
		\begin{minipage}{.3\textwidth}
                        \includegraphics[trim={4.25cm 4.5cm 4cm 4cm},clip,height=22.5mm]{{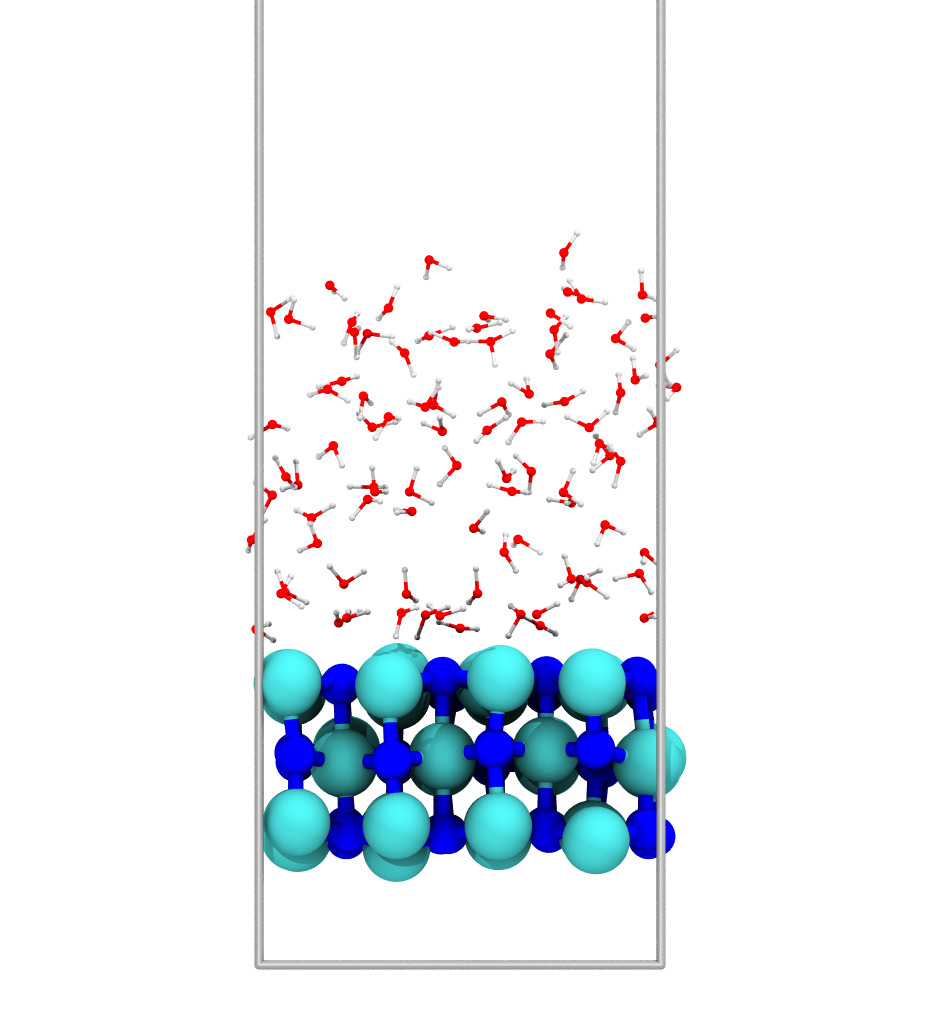}}
                \end{minipage}  \\ \midrule
	\ch{Cl-} / H$_2$O &
	\begin{tabular}[c]{@{}l@{}}N$\mathrm{_{water}}$ = 63\\ N$_{\mathrm{Cl}}$ = 1\\ L = 12.42\,\AA{}\end{tabular} & 
		\begin{minipage}{.3\textwidth}
                        \includegraphics[trim={5cm 5.5cm 3.8cm 4cm},clip,height=23mm]{{Figures/si/box.jpg}}
                \end{minipage} \\ \midrule
        \ch{Na+}--\ch{Cl-} contact ion pair & 
	\begin{tabular}[c]{@{}l@{}}N$\mathrm{_{water}}$ = 62\\ N$_{\mathrm{Na}}$ = 1\\ N$_{\mathrm{Cl}}$ = 1\\ L = 12.42\,\AA{}\end{tabular} &
		\begin{minipage}{.3\textwidth}
                        \includegraphics[trim={4.25cm 4cm 3.8cm 4.7cm},clip,height=23mm]{{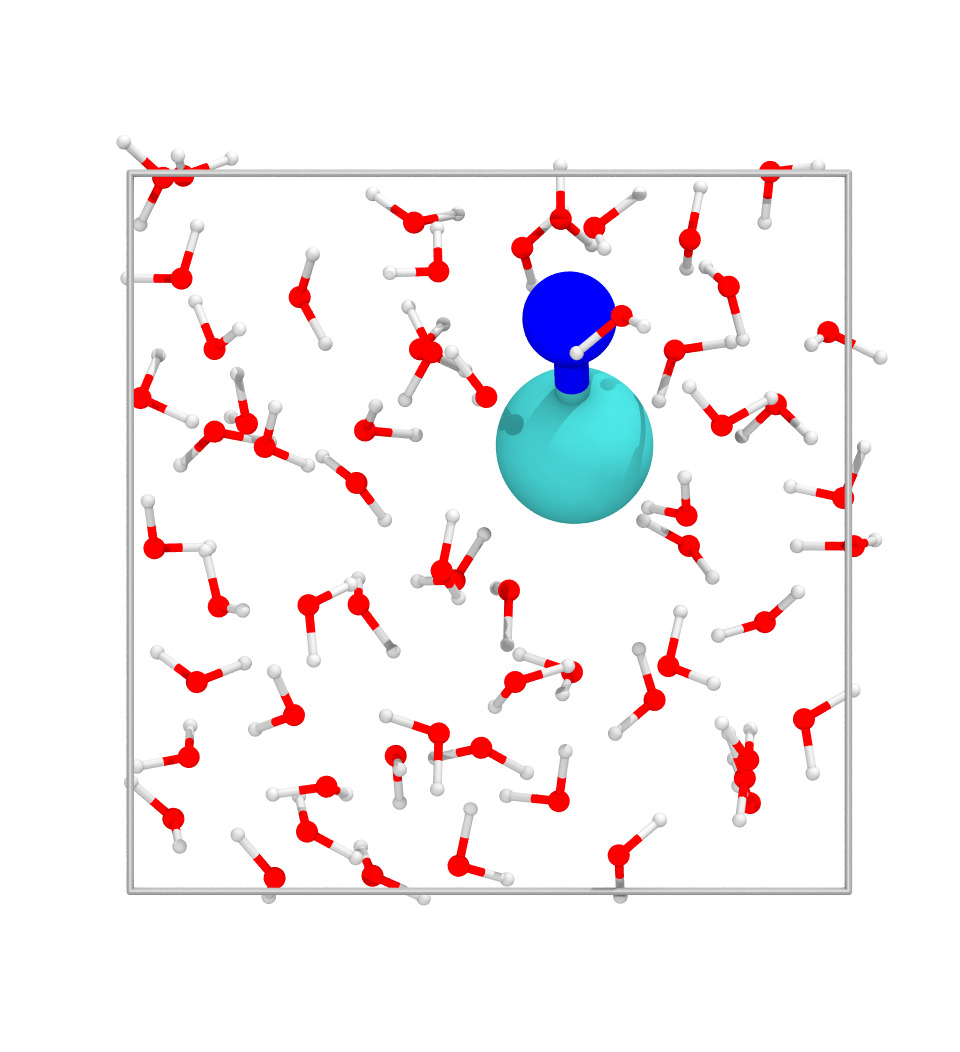}}
                \end{minipage} \\ \midrule
	\ch{Na+}--\ch{Cl-} solvent-separated ion pair
	& \begin{tabular}[c]{@{}l@{}}N$\mathrm{_{water}}$ = 62\\ N$_{\mathrm{Na}}$ = 1\\ N$_{\mathrm{Cl}}$ = 1\\ L  = 12.42\,\AA{}\end{tabular} &
		\begin{minipage}{.3\textwidth}
                        \includegraphics[trim={2cm 3.25cm 2.75cm 4cm},clip,height=21mm]{{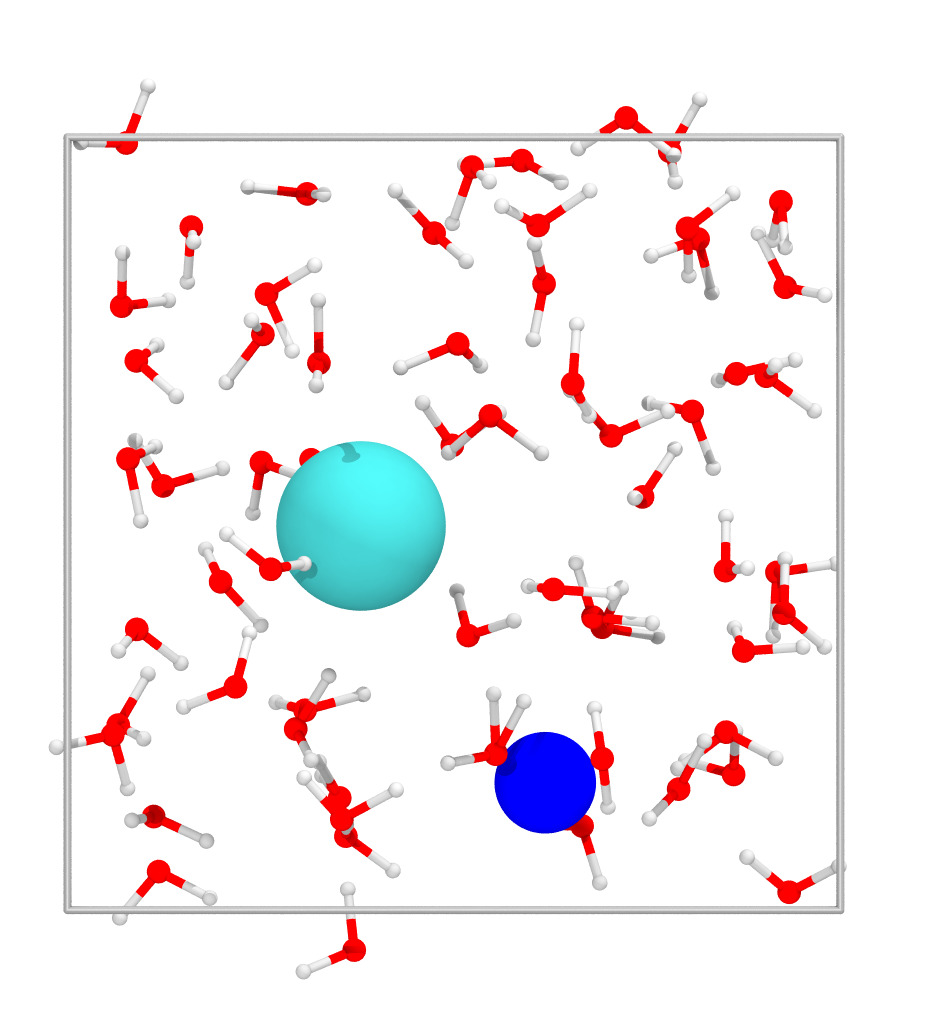}}
                \end{minipage} \\ \midrule
	NaCl / H$_2$O interface &
	\begin{tabular}[c]{@{}l@{}}N$\mathrm{_{water}}$ = 89\\ N$_{\mathrm{Na}}$ = 36\\ N$_{\mathrm{Cl}}$ = 36\\ L(x,y,z) = (16.2, 12.1 58.7)\,\AA{} \end{tabular} & 
		\begin{minipage}{.3\textwidth}
                        \includegraphics[trim={8cm 1.25cm 2.75cm 4cm},clip,height=40mm]{{Figures/si/box.jpg}}
                \end{minipage} \\ \midrule
	$2 \times 2 \times 2$ NaCl nanocrystal
	& \begin{tabular}[c]{@{}l@{}}N$\mathrm{_{water}}$ = 229\\ N$_{\mathrm{Na}}$ = 4\\ N$_{\mathrm{Cl}}$ = 4\\ L = 19.8\,\AA{}\end{tabular} &
		\begin{minipage}{.3\textwidth}
                        \includegraphics[trim={1cm 2.5cm 1.5cm 3cm},clip,height=27mm]{{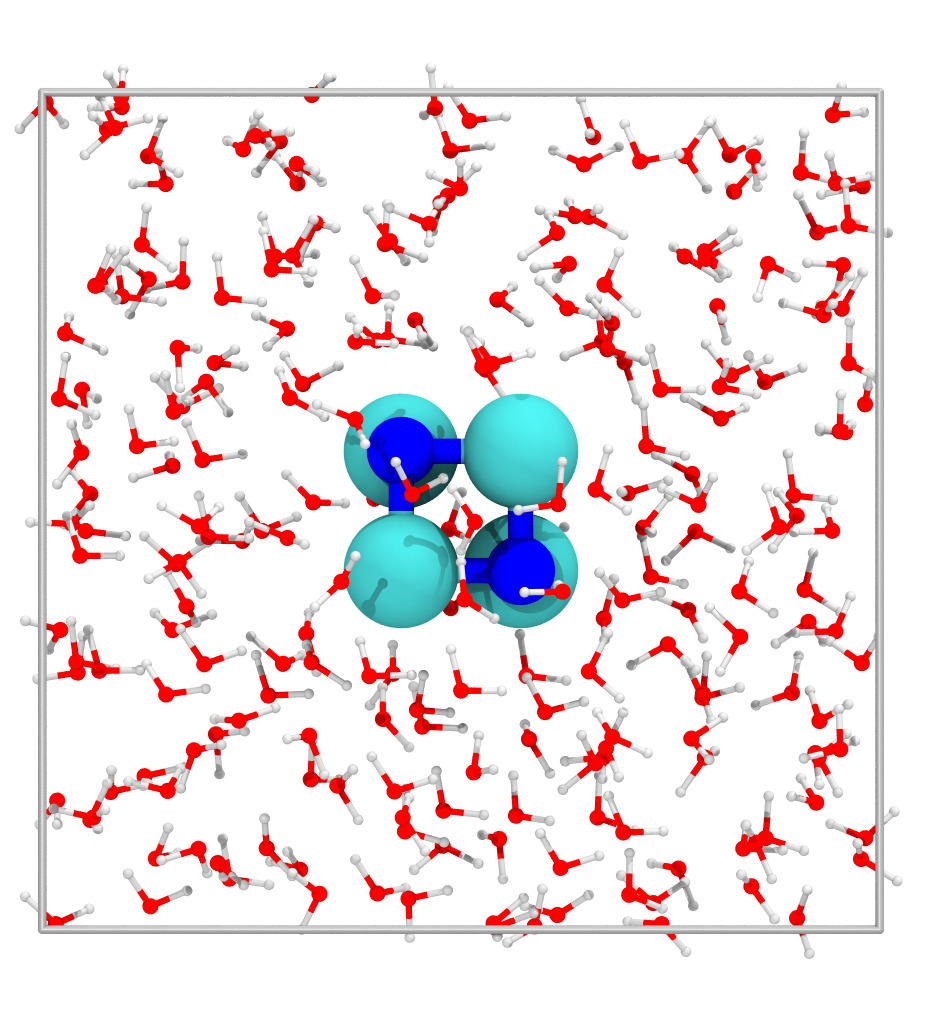}}
                \end{minipage}  \\ \midrule
	$4\times4\times4$ NaCl nanocrystal (2.48 mol/kg)
	& \begin{tabular}[c]{@{}l@{}}N$\mathrm{_{water}}$ = 625\\ N$_{\mathrm{Na}}$ = 32\\ N$_{\mathrm{Cl}}$ = 32\\ L = 27.8\,\AA{}\end{tabular} &
		\begin{minipage}{.3\textwidth}
                        \includegraphics[trim={1.5cm 3.25cm 2cm 3.5cm},clip,height=30mm]{{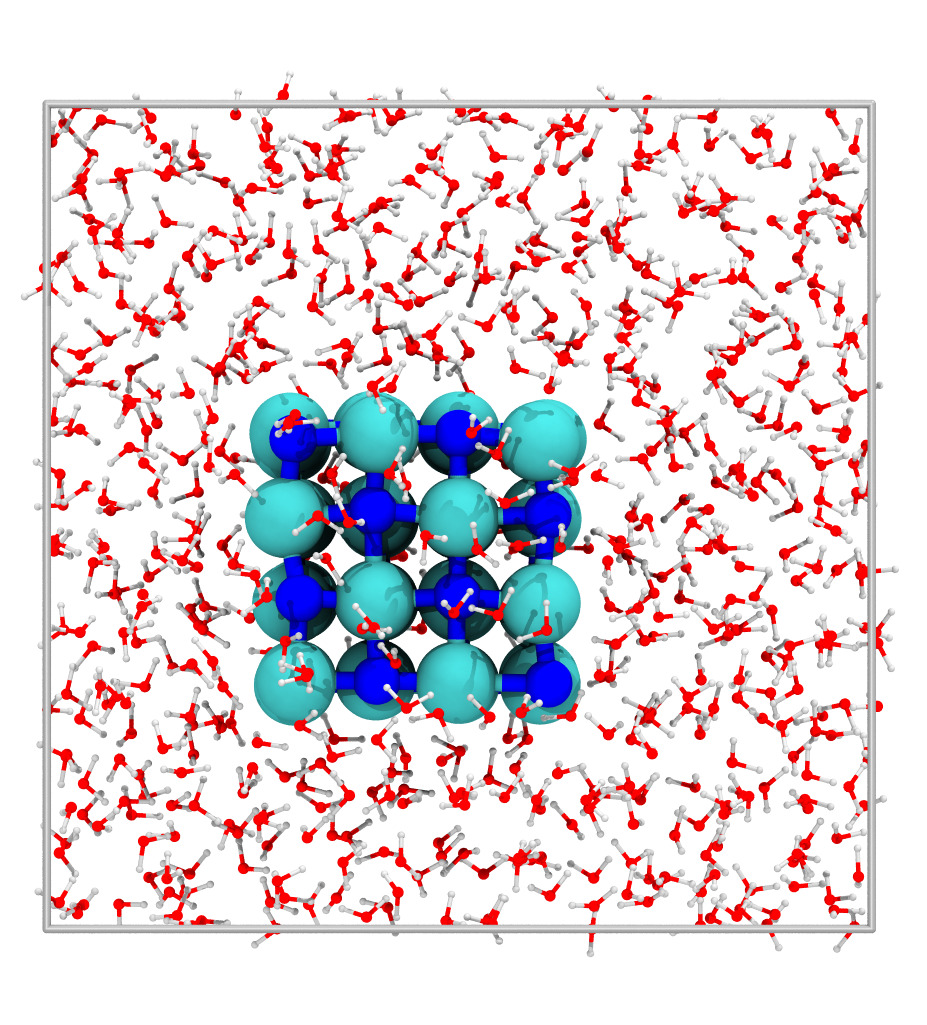}}
                \end{minipage} \\ \midrule
	$4\times4\times4$ NaCl nanocrystal (1.42 mol/kg)
	& \begin{tabular}[c]{@{}l@{}}N$\mathrm{_{water}}$ = 1250\\ N$_{\mathrm{Na}}$ = 32\\ N$_{\mathrm{Cl}}$ = 32\\ L = 34.1\,\AA{} \end{tabular} & 
		\begin{minipage}{.3\textwidth}
                        \includegraphics[trim={3.95cm 4.75cm 3.5cm 4.25cm},clip,height=40mm]{{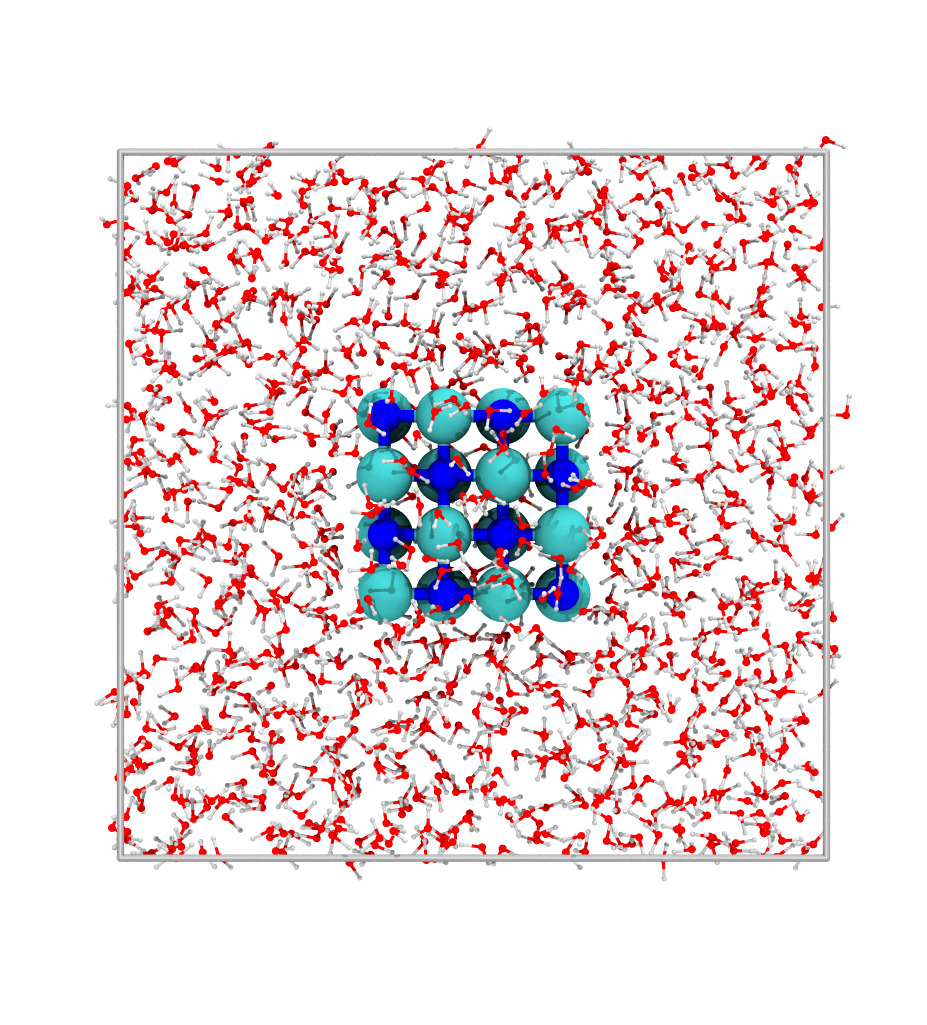}}
                \end{minipage} \\ \midrule 
        $6\times6\times6$ NaCl nanocrystal (5.61 mol/kg)
        & \begin{tabular}[c]{@{}l@{}}N$\mathrm{_{water}}$ = 1068\\ N$_{\mathrm{Na}}$ = 108\\ N$_{\mathrm{Cl}}$ = 108\\ L = 34.1\,\AA{} \end{tabular} &
                \begin{minipage}{.3\textwidth}
                        \includegraphics[trim={4.2cm 4.75cm 3.75cm 5cm},clip,height=40mm]{{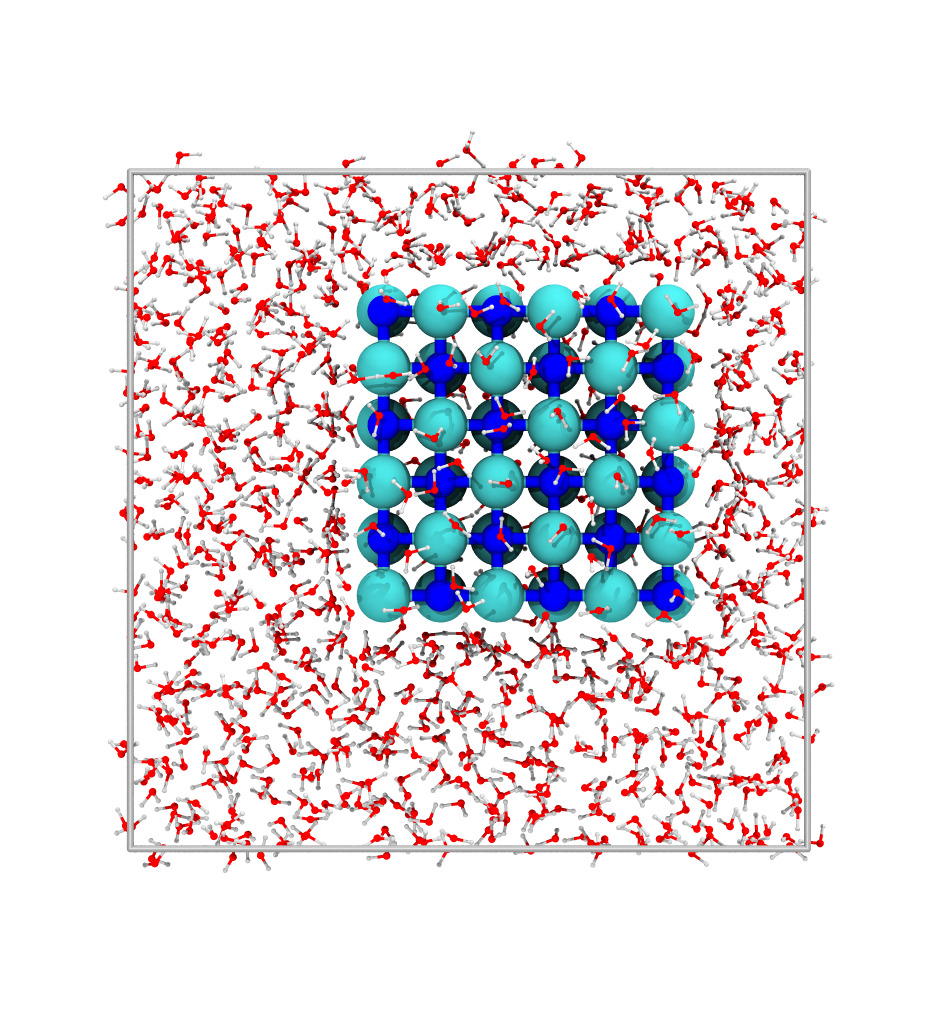}}
                \end{minipage}\\
		\bottomrule
\end{longtable}
\end{spacing}

\begin{table}[h!]
\setlength{\tabcolsep}{12pt}
\caption{Details of simulation lengths performed with AIMD. }
\centering
\begin{tabular}{@{}ll@{}}
\toprule
\multicolumn{2}{c}{\textbf{AIMD}} \\ \midrule
\textbf{System} & \multicolumn{1}{c}{$\mathbf{T_{sim}}$} \\ \midrule
Bulk H2O & 100 ps \\ \midrule
Na / H2O & 80 ps \\ \midrule
Cl / H2O & 70 ps \\ \midrule
NaCl/H2O interface & 70 ps \\ \midrule
Na--Cl contact ion pair & 36 ps \\ \midrule
Na--Cl solvent-separated ion pair & 36 ps \\ \midrule
2 x 2 x 2 NaCl nanocrystal & 25 ps \\ \bottomrule
\end{tabular}
\label{tab:aimd}
\end{table}

\begin{table}[h!]
\setlength{\tabcolsep}{12pt}
\caption{Details of simulation lengths for production runs performed with C-NNP. }
\centering
\begin{tabular}{@{}ll@{}}
\toprule
\multicolumn{2}{c}{\textbf{C-NNP Production }} \\ \midrule
\textbf{System} & \multicolumn{1}{c}{$\mathbf{T_{sim}}$} \\ \midrule
$4 \times 4 \times 4$ NaCl nanocrystal (2.84 mol/kg) & 10 $\times$ \textgreater{}2 ns \\ \midrule
$4 \times 4 \times 4$ NaCl nanocrystal (1.42 mol/kg) & 10 $\times$ \textgreater{}2 ns \\ \midrule
$6 \times 6 \times 6$ NaCl nanocrystal (5.61 mol/kg) & 10 $\times$ 10 - 30 ns \\ \midrule
\end{tabular}
\label{tab:cnnp}
\end{table}

\subsection{Simulation details}
All MD/AIMD simulations were performed using the CP2K/Quickstep code \cite{Kuhne2020/10.1063/5.0007045}. 
AIMD simulations were used to generate reference data for development of the model as shown in Table \ref{tab:aimd}, with pre-equilibrated structures obtained from force field simulation. C-NNP production simulations as shown in Table \ref{tab:cnnp} were initially equillibrated with fixed ions using the C-NNP model. 

\subsubsection{Force field pre-equilibration details}
A 1 ns force-field based MD simulation was first performed on all Na/Cl containing systems from Table \ref{tab:aimd} to obtain a pre-equilibrated initial configuration for subsequent AIMD simulations.
The Born-Mayer-Huggins potential for NaCl \cite{Fumi1964/10.1016/0022-3697(64)90159-3} and TIP3P model for water \cite{Jorgensen1983/10.1063/1.445869} were used and the water/NaCl interactions were described by the parameters given by Lynden-Bell et al. in Ref.~\citenum{Lynden-Bell1997/10.1063/1.474550}.

Additional production simulations as described in Section \ref{sec:FF} were performed using the Joung Cheatham (JC) model for ion-ion interactions, with SPC/E water model and ion-water interactions described by LJ crossed interactions using the Lorentz-Berthelot combination rules. These parameters were taken from Ref. \citenum{Benavides2016/10.1063/1.4943780}.
\subsubsection{AIMD simulation details}
Prior to all AIMD simulations, a short constrained AIMD simulation, keeping ions fixed, was performed on all the systems given in Table \ref{tab:aimd} containing ions, to obtain a well-equillibrated water structure around the Na and Cl ions.
These were used as starting configurations for simulations in NVT ensemble within the generalised gradient approximation using the revPBE functional \cite{Zhang1998/10.1103/PhysRevLett.80.890} with Grimmes' dispersion correction \cite{Grimme2010/10.1063/1.3382344}.
The electronic density was partitioned into core and valence contributions, with core electrons described using the norm-conserving Goedecker, Teter and Hutter (GTH) pseudopotentials \cite{Goedecker1996/10.1103/PhysRevB.54.1703}.
Na 2s and 2p electrons were also treated explicitly given the well-known issue of non-linear core-valence exchange/correlation.
Valence electrons were described using the MOLOPT TZV2P basis set \cite{VandeVondele2007/10.1063/1.2770708}.
Stochastic errors for Na forces due to insufficient basis set convergence were observed in additional DFT convergence tests for the cutoff of the plane-wave representation of the charge density.
Therefore a plane wave cutoff of 1200 Ry was used to obtain accurate forces.
Deuterium masses were used in lieu of hydrogen, allowing a 1 fs timestep and reducing errors due to neglecting nuclear quantum effects through classical propagation of the nuclei.
Simulations were performed at 300 K, maintained using the CSVR thermostat. \cite{Bussi2007/10.1063/1.2408420}.
All AIMD simulations were at least 40 ps long (except for $2 \times 2 \times 2$ NaCl nanocrystal which was 25 ps due to computational cost.) and can be found in Table \ref{tab:aimd}.

\section{Analysis Details}
\subsection{Steinhardt order parameters}
The Steinhardt order parameter q8 is a particular case of a larger set of bond order parameters based on spherical harmonics.~\cite{Steinhardt1983/10.1103/PhysRevB.28.784}
To distinguish between solid and liquid-like ions we first compute the complex vector $q_8(i)$ for each particle $i$, where $N_b(i)$ is the number of nearest neighbours of particle $i$, and $m$ runs from -8 to +8.
\begin{equation}
	q_{8,m}(i) = \frac{1}{N_b(i)}\sum_j^{N_b(i)}Y_{8,m}(\mathbf{r}_{i,j})
\end{equation}
$Y_{8,m}$ are the spherical harmonics and $\mathbf{r}_{i,j}$ connects particles $i$ and $j$.
Following Lechner et al. \cite{Lechner2008/10.1063/1.2977970} we then compute an averaged form of this local bond order parameter by averaging the local $q_{8,m}(i)$ vectors over the particle $i$ and its surroundings:
\begin{equation}
	\bar{q}_8(i) = \sqrt{\frac{4\pi}{2l+1}\sum_{m=-l}^l|{q_{8,m}(i)}|^2}
\end{equation}
where
\begin{equation}
	\bar{q}_{8,m}(i) = \frac{1}{\tilde{N}_b(i)}\sum_{k=0}^{\tilde{N}_b(i)}Y_{8,m}(k)
\end{equation}
and $\tilde{N_b(i)}$ are the neighbours of particle $i$

\begin{figure*}[ht]
    \centering
    \includegraphics{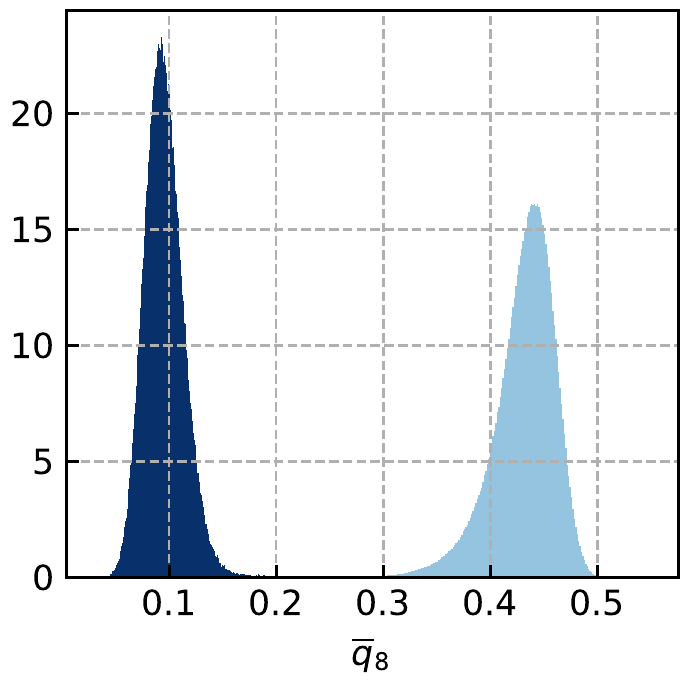}

    \caption{$\bar{q}_8$ distributions for ideal cases of pristine NaCl lattcice (light blue) and solvated Na/Cl ions (dark blue).
    }
    \label{fig:q8-distribution}
\end{figure*}

Computation of these bond order parameters was performed using the Freud \cite{Ramasubramani2020/10.1016/j.cpc.2020.107275} python package.

\subsection{Coordination Numbers and clusters}
\begin{figure*}[ht]
    \centering
    \includegraphics{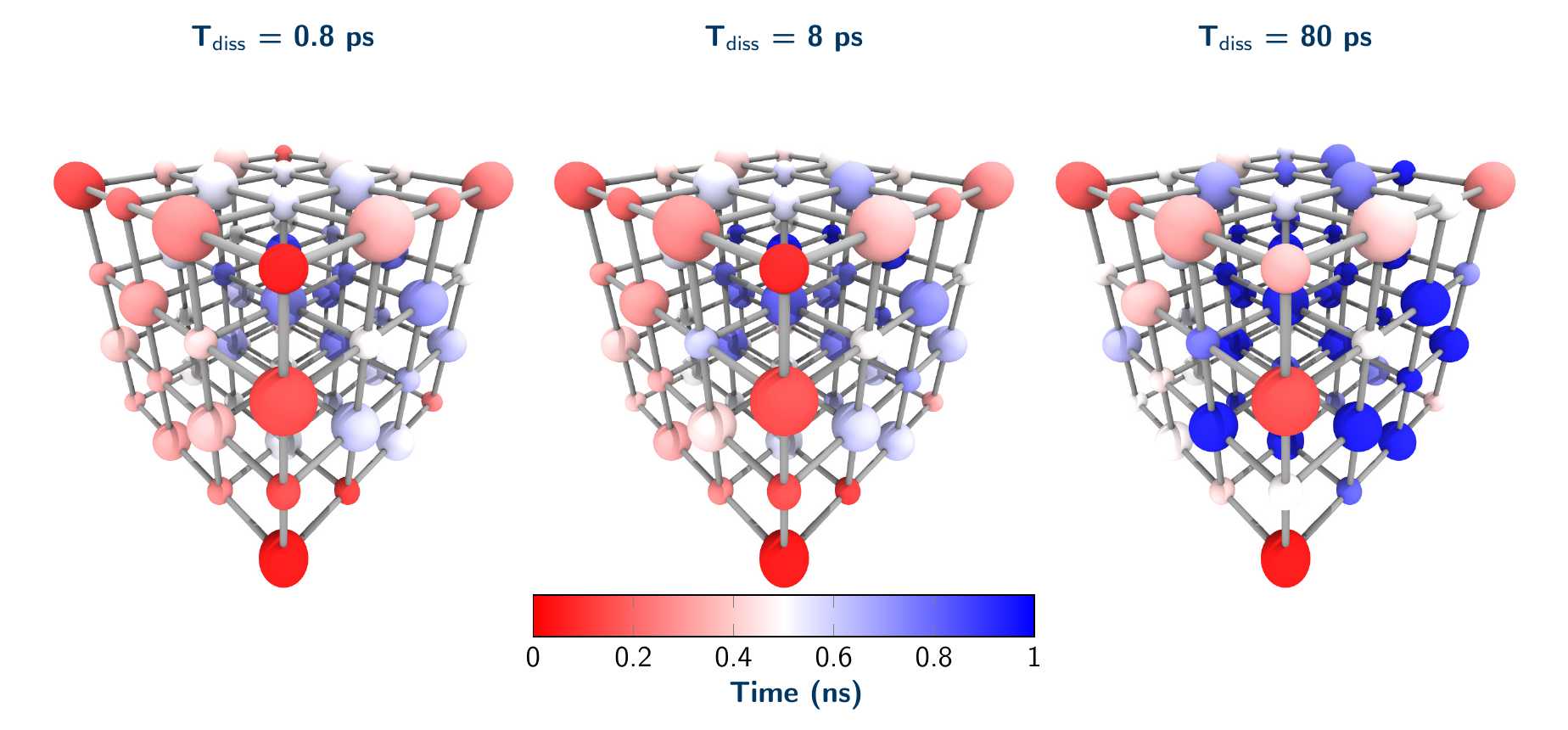}

    \caption{Sensitivity of dissolution behaviour of ions based on definition of time spent with CN$<$1.
    }
    \label{fig:rubics}
\end{figure*}

Computation of all coordination numbers and clusters were performed using Plumed \cite{Bonomi2019/10.1038/s41592-019-0506-8}, and analysis carried out using Python.
The CN of an atom $i$ with a set of atoms in $A$ is defined by:
\begin{equation}
    \text{CN}_i=\sum_{j\in A} s_{ij}
\end{equation}
where $s_{ij}$ is a smooth switching function with a range of [0,1]:
\begin{equation}
    s_{ij}=\left(
    \frac{1-\left(\frac{r_{ij}}{r_0}\right)^6}
         {1-\left(\frac{r_{ij}}{r_0}\right)^{12}}
    \right)
\end{equation}
$r_{ij}$ is the interatomic distance between atom $i$ and atom $j$.
$r_0$ for Na-O (2.80 \AA{}) and Cl-H (2.60 \AA{}) were obtained from literature values \cite{Liu2011/10.1039/C1CP21077G}, while $r_0$ for Na-Cl was taken such that the CNs of the pristine lattice coincided with the integral up to the first minimum of the Na-Cl RDF (3.5 \AA{}).
An ion was defined as dissolved if it has a Na-O/Cl-H CN less than 1 for more than 8 ps.
We show that this definition is not sensitive to changes in the time chosen in Figure \ref{fig:rubics}.
The cluster size distributions were computed by detecting coordinated ions using a threshold of a ion-ion CN $>$ 2, and then using these atoms as input for a depth first search graph reduction algorithm to identify the sets of ions clustered together.

To compute the surface area to volume ratio of the dissolving crystal, a DFS algorithm was implemented using the Freud Python package \cite{Ramasubramani2020/10.1016/j.cpc.2020.107275} in order to identify the ions in the largest cluster at a given time.
The convex hull of this set of points was then obtained (provided there were $>3 $ ions in the cluster), and the surface area to volume ratio of the hull was computed.

\section{Additional Results}\label{sec:FF}

\subsection{MLP}
\begin{figure*}[h]
    \centering
    \includegraphics{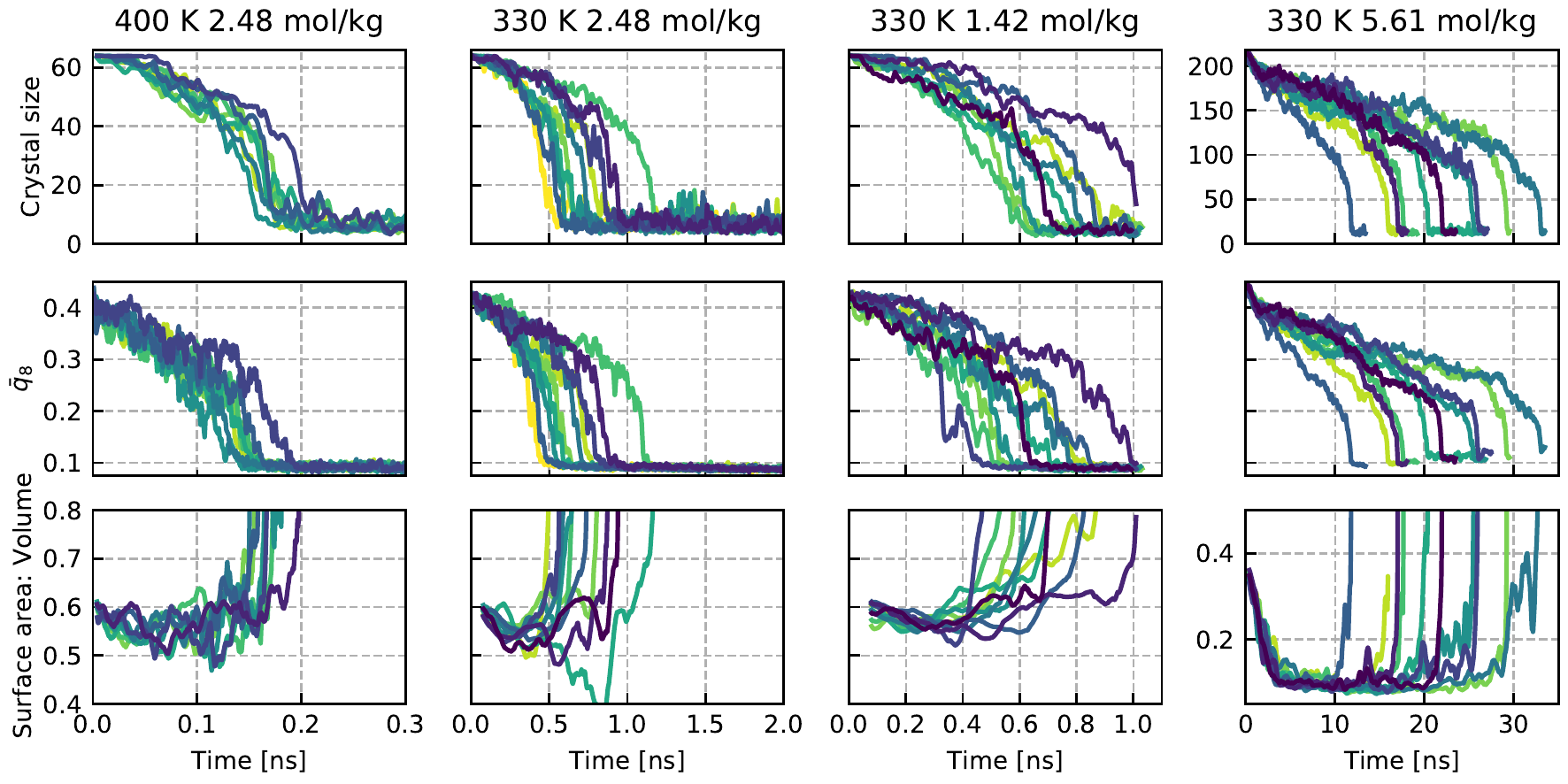}

	\caption{Crystal size, $\bar{q}_8$ and surface area:volume ratio for ten simulated trajectories at 400 K (2.48 mol/kg) and 330 K (2.48 mol/kg, 1.42 mol/kg and 5.61 mol/kg).
    }
    \label{fig:400K}
\end{figure*}

Results from identical simulations as described above for the C-NNP at 2.48 mol/kg NaCl concentration at 400 K are shown in Figure \ref{fig:400K}. 
The additional results for the 1.42 mol/kg and 2.48 mol/kg trajectories at 330 K are also shown in Figure \ref{fig:400K}.

\subsection{FF}
\begin{figure*}[h!]
    \centering
    \includegraphics{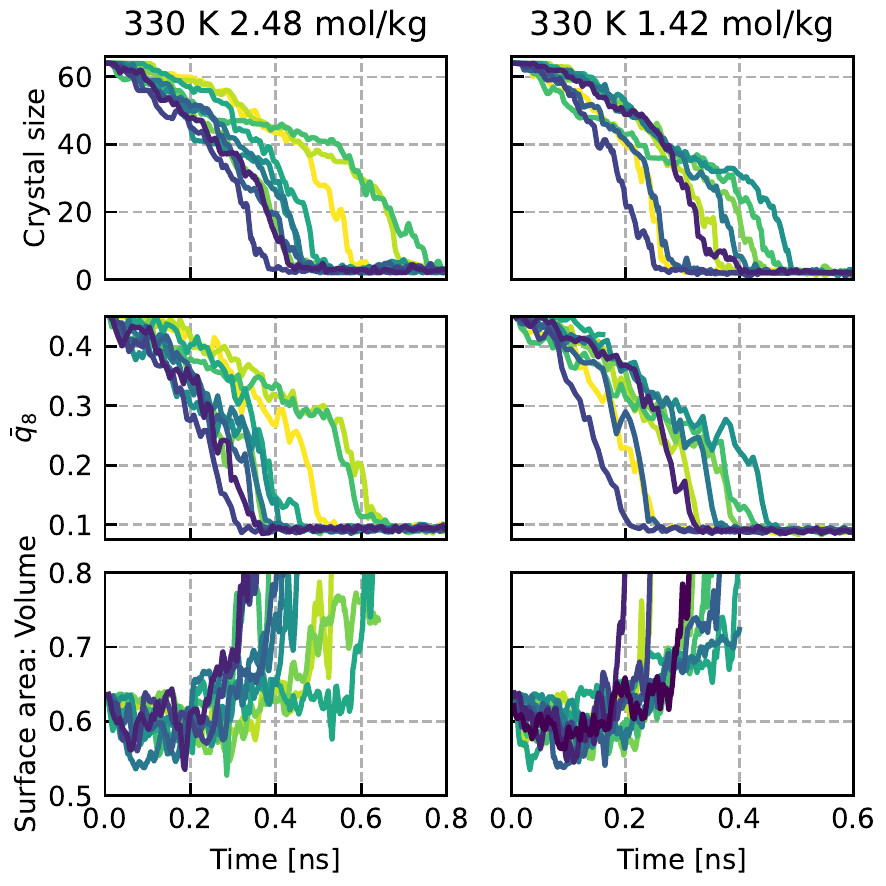}

	\caption{Crystal size, $\bar{q}_8$ and surface area:volume ratio for ten simulated trajectories with JC/SPC/E at 330 K (2.48 mol/kg and 1.42 mol/kg).
    }
    \label{fig:FF}
\end{figure*}

Results from identical simulations as described above for the C-NNP at 1.42 mol/kg and 2.48 mol/kg at 330 K are shown in Figure \ref{fig:FF} for a JC/SCP/E force-field model. 

\FloatBarrier
%
\section*{References}

%